\documentclass[preprint,11pt]{aastex}
\usepackage{amssymb,amsmath}
\usepackage{color,nicefrac}
\def\xc{\emph{X Comae}}
\def\coma{\emph{Coma}}
\def\hst{\it HST\rm}

\def\rgs{\emph{RGS}}
\def\nodata{---}

\def\oi{O~I}
\def\oii{O~II}
\def\ovi{O~VI}
\def\ovii{O~VII}
\def\oviii{O~VIII}
\def\nvi{N~VI}

\def\cvi{C~VI}

\def\cv{C~V}

\def\nvii{N~VII}
\def\neii{Ne~II}
\def\neiii{Ne~III}
\def\neix{Ne~IX}
\def\nex{Ne~X}
\def\pg{PG~1116+215}
\def\planck{\emph{Planck}}
\def\pspc{\emph{PSPC}}
\def\chandra{\it Chandra\rm}
\def\xmm{\emph{XMM-Newton}}
\def\rosat{\emph{ROSAT}}
\def\euve{\emph{EUVE}}
\def\suzaku{\emph{Suzaku}}
\def\sdss{\emph{SDSS}}
\begin{document}
\title{Characterization of the warm--hot intergalactic medium near the \coma\ cluster
through high--resolution spectroscopy of \xc}

\author{M. Bonamente\altaffilmark{1,2}, 
%J. Nevalainen\altaffilmark{3}, 
J. Ahoranta\altaffilmark{3},
E. Tilton\altaffilmark{4}, E. Tempel\altaffilmark{5,6},
%P. Hein\"am\"aki\altaffilmark{7} 
and A.~Morandi\altaffilmark{1}
% and T. Fang\altaffilmark{5}. 
}

\altaffiltext{1}{Department of Physics, University of Alabama in Huntsville,
Huntsville, Al}
\altaffiltext{2}{NASA National Space Science and Technology Center, Huntsville, Al}
\altaffiltext{3}{University of Helsinki, FI-00014 Helsinki, Finland}
\altaffiltext{4}{CASA, Department of Astrophysical \& Planetary Sciences, University of Colorado, Boulder, CO 80309}
\altaffiltext{5}{Tartu Observatory, Observatooriumi 1, 61602 T{\~{o}}ravere, Estonia}
\altaffiltext{6}{Leibniz-Institut f\"{u}r Astrophysik Potsdam (AIP), An der Sternwarte 16, D-14482 Potsdam, Germany}
%\altaffiltext{7}{Tuorla Observatory, V\"ais\"al\"antie 20, 21500, Piikki\"o, Finland}
%\altaffiltext{5}{Department of Astronomy and Institute for Theoretical Physics and Astrophysics, Xiamen University, Xiamen, Fujian, China }

\begin{abstract}
We have analyzed all available archival \xmm\ observations of \xc, a bright X--ray quasar behind the
\coma\ cluster, to study the properties of the warm--hot intergalactic medium
in the vicinity of the nearest massive galaxy cluster.
%The new analysis is prompted by the availability of additional
%\xmm\ data that had not been previously analyzed, and by
%the desire to determine the effect of new calibration information and
%methods of analysis compared with the earlier analysis.
%
 The \rgs\ observations confirm the possible presence of a \neix\ K$\alpha$
absorption line at the redshift of \coma, although with a limited statistical
significance. This analysis is therefore in
 line with the earlier analysis 
by \cite{takei2007} based on a sub--set of these data. 
Its large column density and optical depth, however, point to implausible 
conditions for the absorbing medium, thereby casting serious doubts to its reality.
%Our analysis of \sdss\ data was inconclusive in finding a
%suitable large--scale filament of galaxies along the sightline, and
%showed that no individual nearby  galaxy can produce this signal.
\chandra\ has never observed \xc\ and therefore cannot
provide additional information on this source. 
We combine upper limits to the presence of other X--ray absorption lines
(notably from \ovii\ and \oviii) at
the redshift of \coma\ with positive measurements of the soft excess emission
from \coma\ measured by \rosat\ \citep{bonamente2003}. The combination of 
emission from warm--hot gas at $kT \sim \nicefrac{1}{4}$~keV and upper limits
from absorption lines provide useful constraints on the density
and the sightline length of the putative WHIM towards \coma.
We conclude that the putative warm--hot medium towards \coma\ 
is consistent with expected properties, with a baryon overdensity $\delta_b \geq 10$ and a sightline extent of order of tens of Mpc.
\end{abstract}


\section{The search for missing baryons}

The diffuse intergalatic medium contains the majority of the universe's baryons
at all redshifts \citep[][and references therein]{shull2012}.
At high redshift, the bulk
 of this mass is in the photoionized phase that gives rise to the Lyman-$\alpha$ forest \citep[e.g.,][]{penton2000}.
At lower redshift, numerical simulations predict that 
 a diffuse warm--hot intergalactic medium (WHIM) at temperatures
 $\log T(K)=5-7$ contain approximately 50\% of 
the baryons in the universe \citep[e.g.][]{cen1999,dave2001}.
Far ultraviolet (FUV) absorption-line spectroscopy, probing the lower--temperature 
phase of the WHIM ($\log T(K)=5-6$) with transitions from species such as \ovi, 
has so far detected only a fraction of the expected baryons \citep[e.g.,][]{danforth2008,tripp2008,tilton2012,shull2012,
danforth2016}.
The remaining low--redshift baryons are likely to be found at higher temperature ($\log T(K)=6-7$)
where X--ray observations can be used to detect their high--ionization absorption lines
(e.g., \ovii\ and \oviii\ K$\alpha$).

Of particular interest are sightlines near massive galaxy groups and clusters \citep[e.g.,][]{werner2008}.
During the formation of galaxy groups, baryons accrete from large--scale
filamentary structures towards their gravitational potential
and the infall heats them by  loss of gravitational energy. A portion of these
baryons form the virialized, high--temperature ($\log T(K) \geq 7$) 
and relatively high--density ($\geq 10^{-5}$ cm$^{-3}$) intra--cluster medium.
A significant amount of baryons, however, is expected
to remain at sub--virial temperatures near the outskirts
of clusters. These sub--virial baryons
are harder to observe because of their lower temperature and density, and
 may provide a significant reservoir of low--density WHIM baryons.

\section{The  sightline towards \xc\ and the \coma\ cluster}
\label{sec:los}
In this paper we examine the \xmm\ high--resolution \rgs\ spectra of
\xc, a Seyfert galaxy at redshift $z=0.091\pm0.001$ \citep{Branduardi1985},
in the background of the massive, low--redshift \coma\ cluster
at $z=0.023$ \citep{struble1999}.
\xc\ is the brightest X--ray
source behind the \coma\ cluster, at a projected distance of approximately
28.4' (or 790~kpc) and located to the north of the cluster core. 
The projected position of \xc\ falls well within the \coma\ virial radius \citep[e.g., $r_{200}=1.99\pm0.21 h^{-1}$ Mpc,][]{kubo2007}.
 We use a distance scale of 27.8 kpc~arcmin$^{-1}$
at the redshift of the \coma\ cluster, for $h=70$ km~s$^{-1}$~Mpc$^{-1}$, $\Omega_m=0.3$ and a
flat universe. Our goal is to examine the presence of absorption lines from the putative WHIM
at the redshift of \coma\ with the analysis of all available optical and \rgs\ data. 
The Coma cluster has  a very diffuse halo of hot X--ray emitting intra--cluster medium (ICM).
The hot ICM  has an integrated
temperature of $8.2$~keV \citep{hughes1993} and a temperature
of $6-8$ keV in the central 20' region \citep{arnaud2001}. 
\cite{simionescu2013} measures a radial profile of the hot ICM temperature from \suzaku,
with a temperature of $5-10$~keV at radii 20-40' (the
projected distance of \xc).
In peripheral regions at $\leq 1$ degree from the center, \cite{finoguenov2003} measured
temperatures varying from 3 to $\geq 10$~keV
with \xmm. 
\cite{takei2007} also reported a temperature of $3.75\pm^{0.32}_{0.50}$~keV for the
region in the foreground of \xc\ also with \xmm, using a sub--set of 
the observations presented in this paper. 

The \coma\ cluster also features a large--scale 
halo of warm gas at sub--virial temperatures 
that was discovered from \rosat\ and \euve\ \citep{lieu1996b,bonamente2003,bonamente2009}. 
Pointed \rosat\ observations that
cover the entire \coma\ cluster measured a temperature
of $kT \simeq\ \nicefrac{1}{4}$~keV for the warm gas, in addition to the well--known
hot intra--cluster medium. The warm gas temperature measured by \rosat\
 varies slightly with radius and azimuth with a typical uncertainty of order $0.05$~keV and
an abundance of approximately $\leq 0.1$~Solar (see Table~3 in \citealt{bonamente2003}). 
The analysis of \xmm\ CCD imaging spectrometer data
by \cite{finoguenov2003} also detected soft X--ray emission at $kT \simeq 0.25$~keV
in several peripheral regions of \coma\ (fields named \coma-0, 3, 7 11 and 13). 
The analysis of the \suzaku\ data of field \coma-11 by \cite{takei2008}, however, failed
to detect the presence of oxygen emission lines from the WHIM.
A possibility for the discrepancy between the \cite{finoguenov2003} and \cite{takei2008} results
on \coma-11 is
the presence of solar wind charge-exchange (SWCX) radiation.
This soft X--ray radiation from the local Solar enviroment 
 may be mis--identified as
WHIM emission in low--resolution CCD spectra, if not properly subtracted. 
The \suzaku\ data does not address the presence of the
soft excess continuum detected with \rosat\ because \suzaku\ does not cover the 
 $\nicefrac{1}{4}$~keV energy band available with \rosat.
The \rosat\ data of \cite{bonamente2003, bonamente2009}
are not affected by SWCX radiation problems
because we used \emph{in situ} simultaneous background, thanks to the large field of
view of the \emph{Position--Sensitive Proportional Counter} (\pspc) detector of \rosat.

\cite{takei2007} also analyzed a sub--set of the \xmm\ data presented in this paper,
in search of both emission and absorption lines 
from the putative WHIM near \coma. 
Their
 analysis
of the higher--resolution \rgs\ spectra of \xc\
resulted in the detection of absorption lines
at the wavelengths of the redshifted  \neix\ line (2.3 $\sigma$ confidence)
 and \oviii\ line (1.9 $\sigma$).
In addition, from the analysis of low--resolution \emph{EPIC}
CCD spectra they report
a 3.4$\sigma$ detection of excess flux that corresponds to a redshifted \neix\ 
K$\alpha$ line.
In this paper we do not re--analyze the
\xmm\ \emph{EPIC} data towards \xc, since our aim is to study lines from the WHIM
and the spectral resolution of \emph{EPIC} is insufficient to separate accurately
the strong Galactic $z=0$ lines from those at the redshift of \coma.
Moreover, the \rosat\ data we use to constrain the continuum emission from the 
WHIM have a number of advantages over the available \emph{EPIC} data,
including a full azimuthal coverage, low background and insensitivity to
local particle background. More importantly,  \rosat\ covers directly the $\nicefrac{1}{4}$~keV
band where the WHIM emission is expected to be strongest, and the effective
area of \rosat\ is well calibrated at these energies \citep[e.g.][]{snowden1994}. 
At the end of Section~\ref{sec:WHIM}
we compare our constraints on the properties of the WHIM based on our
analysis to the \cite{takei2007}
detection of \neix\ emission.
% and conclude that the two results are in agreement.

This soft X--ray emission around \coma\ is part of the
soft excess phenomenon detected in a number of low--redshift clusters, consisting
of substantial amounts of soft X--ray emission in excess of the amount expected
as the low--energy tail of the hot ICM \citep[e.g.][]{bonamente2002}.
A possible explanation for the soft excess is precisely the presence of sub--virial
gas near the cluster outskirts, likely converging towards the massive gravitational
potential from large--scale filaments. 
As WHIM filaments intersect galaxy clusters, we expect that the WHIM becomes
more dense and therefore more readily observable towards clusters.
This re--analysis of \xc\ in search for absorption lines
is therefore useful to constrain
the presence of $\sim \nicefrac{1}{4}$~keV WHIM gas at the redshift of  \coma.
%Despite \xc's X--ray flux and unique location, \chandra\ never observed this source. 
%Observations with \chandra\ would be beneficial to provide an independent assessment of
%the presence of WHIM absorption lines. 

%In addition, \hst\ only has a snapshot observation (750~s)
%of \xc\ with FOS/G130H that does not permit the analysis of associated
%FUV lines such as \hi\ Lyman-$\alpha$ or from \civ\ and \nv. 
%Detections of FUV absorption lines would suggest multitemperature structure along the line of sight 
%and would be beneficial to further understand the WHIM near galaxy clusters \citep[e.g.,][]{emerick2015}.

\begin{table}
\centering
\caption{Log of observations \label{tab:log}}
\begin{tabular}{lcccc}
\hline
Obs. ID & Start Date &  Exp. Time & Clean Exp. Time \\
\hline
0204040101 & 2004-06-06 & 101655.8 & 51698.2 \\
0204040201 & 2004-06-18 & 101857.4 & 37854.3\\
0204040301 & 2004-07-12 & 99486.3  & 26708.7 \\
0304320201 & 2005-06-28 & 80647.4  & 40404.7 \\
030432030 & 2005-06-27 &  55227.5  & 13762.7 \\
0304320801 & 2006-06-06 & 63751.9  & 16946.5 \\
\hline
Total   &    & 502626.3 & 187375.4 \\
\hline
\end{tabular}
\end{table}
%To date, there have been only a handful
%of reported detections of X--ray lines from the WHIM, typically from \ovii\ and \oviii.
%These detections include absorption features toward the targets
% H~2356-309 \citep{buote2009,fang2010}, PKS~2155-304 \citep{fang2002,fang2007,yao2009},
%Mkn~421 \citep{nicastro2005,rasmussen2007,yao2012}, Mkn~501 \citep{ren2014} and
%1ES~1553+113 \citep{nicastro2013}. Given the limited statistical significance of
%all X--ray lines detected to date, it is important to investigate additional
%sightlines and understand the correlation between UV and X--ray absorption lines.

%\begin{figure}
%\centering
%\caption{Light curve of observation 0204040101. The
%level of the quiescent background is shown as the dashed line in the inset,
%which shows a zoom--in of the initial portion of the light curve.
%\label{fig:flare}}
%\includegraphics[width=6in]{lcurve.pdf}
%\end{figure}

\section{\xmm\ observations of \xc\ and data analysis}
\label{sec:data}
\xmm\ observed \xc\ in six separate observations 
for a total of approximately 500~ksec (Table~\ref{tab:log}).
The reduction of the RGS data follow the standard \emph{rgsproc} pipeline
with the \emph{SAS} software, as described in \cite{bonamente2016}.
One of the key steps in the analysis of these data is filtering times
of high background, which are quite common for \xmm.
This step is achived by choosing a level for the quiescent
background rate from the analysis of lightcurves
of the standard \rgs\ spectral extraction regions.
In our analysis we chose a quiescent background level
of $0.05$ counts~s$^{-1}$.
%Fig.~\ref{fig:flare} shows the light curve of the background
%in one of the six exposures. 
There are clear enhancements (flares) 
in the background level that need to be filtered out to 
improve the quality of the resulting source spectrum.
All of the six observations of \xc\ are significantly affected
by background flares.
% in ways similar to those of Fig.~\ref{fig:flare}. 
As a result, the clean exposure time is
reduced to a fraction of the total exposure time (Table~\ref{tab:log}).
In Section~\ref{sec:systematics}
we discuss the impact of this choice by using a less conservative background
level.

Each observation is reduced individually to apply the relevant 
calibration information. For each observation we generate a spectrum
with its background and response function, and combine all
six spectra into one spectrum with the \emph{rgscombine} 
tool. The spectrum is rebinned by a factor of 2 over the initial
resolution of 10~m\AA\ to a bin size of 20~m\AA. This bin size
oversamples the approximate resolution of the 
\xmm\ \rgs\ instrument by a factor of
approximately three, therefore optimizing the resolving power
and the ability to measure narrow line features.
%In Sec.~\ref{sec:systematics} we
%address the effect of our choice of bin size.

\begin{figure}
\includegraphics[width=6in]{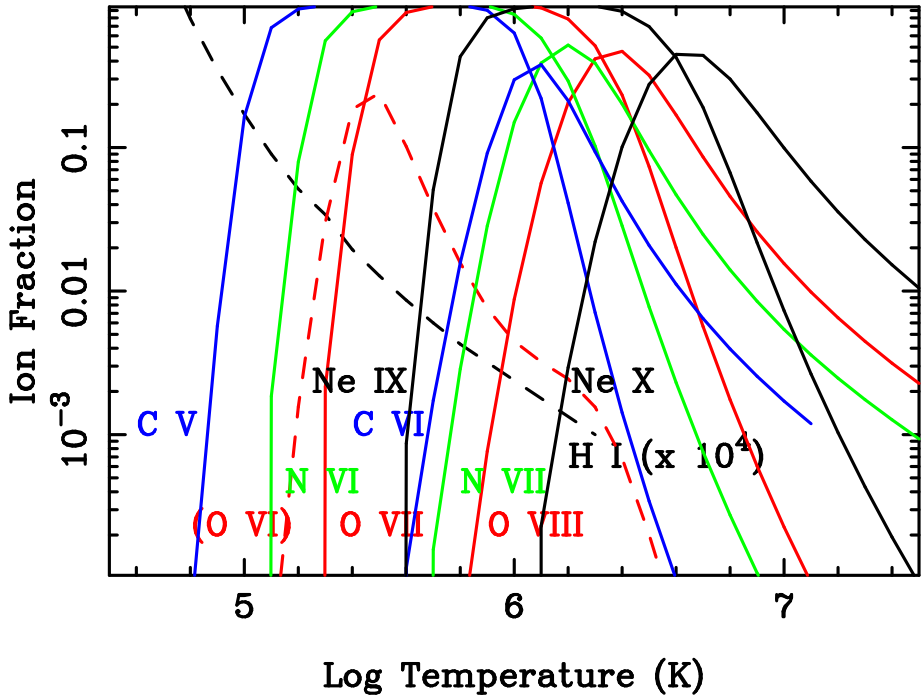}
\caption{Fractions of selected ions with prominent X--ray lines,
 as function of temperature in ionization equilibrium \citep{mazzotta1998}. \label{fig:ions}}
\end{figure}

The goal of this analysis is to place constraints on the
presence of WHIM absorption lines at the redshift of \coma.
In the temperature range $\log T(K) = 6-7$ and assuming
collisional ionization equilibrium, the 
most abundant ions are expected to be the $H$-like and
$He$--like ions of neon, oxygen, nitrogen and carbon, as shown in Figure~\ref{fig:ions} and 
Table~\ref{tab:ions}
(calculations are from \citealt{mazzotta1998} and \citealt{gnat2007}
and atomic data of the lines from \citealt{verner1996}).
For example, if the WHIM is at $0.25\pm0.05$~keV 
(i.e., $\log T(K) \simeq 6.45\pm0.10$), consistent with
the temperature
measured by \cite{bonamente2003},
we expect the K$\alpha$ lines of \oviii\ and \neix\ 
to be particularly prominent.

The analysis of the coadded first--order spectra (one for RGS1 and one for RGS2)  
is performed in \emph{SPEX} \citep{kaastra1996} using the \emph{Cash} statistic
(also known as $\mathcal{C}$ statistic or \emph{cstat}, from \citealt{cash1979}) 
as the fit statistic. In fact, in most 20--m\AA\ bins the number of counts is
$\leq 20$, especially at wavelengths $\lambda \geq 20$~\AA, and the Gaussian approximation
to the Poisson counting statistic would not be accurate. 
The background subtraction in the spectra analysis is 
achieved with the use of a local background that is obtained, from the same observations, as part of the
standard data reduction process. In \emph{SPEX}, the Poisson--based \emph{Cash} statistic
for background--subtracted spectra is implemented in the following way: 
the observed background--subtracted
spectrum $D$, the measured background $B$ and the model $M$ are combined so that
$\hat{D} = D+B$ is the data and $\hat{M}=M+B$ is the model used in the
calculation of the \emph{Cash} statistic (Jelle Kaastra, private communication).
This method enables the use of the Poisson distribution for the \emph{Cash} statistic 
even in the presence of a subtracted background.
For the spectral region near the
\neix\ and \nex\ lines we have successfully checked that
the $\chi^2_{\text{min}}$ statistic gives results that are consistent with the \emph{Cash} statistic (see
Section~\ref{sec:systematics}).

To ensure an accurate modelling of the source's continuum, we model narrow
wavelength ranges in the neighborhood of each line of interest.
We use the physically--motivated
 \texttt{slab} model
that calculates all transmission properties of the selected lines
including their optical depth and a consistent calculation
of the resulting column density based on each line's curve of growth.
One of the model
parameters is the line broadening parameter $b$, which includes all sources of broadening. As shown
in \cite{bonamente2016}, a large value of $b$ keeps
absorption lines in the linear portion of their curves of
growth for larger values of the column density. Since we seek upper limits to
the non--detection of absorption lines, we used a purely thermal
broadening velocity of $b=50-100$~km~s$^{-1}$, which approximately corresponds to
the thermal velocity of neon, oxygen and nitrogen
at $\log T(K) \simeq 6.0-6.5$.
Broadening of the lines cannot be directly constrained using the
available data.
The presence of non--thermal broadening
would reduce the optical depth of the line, leading to lower upper limits
to the column density.
In Table~\ref{tab:spex} we report the parameters
of the \texttt{slab} SPEX fit.
Although the \rgs\ instrument covers the band where \nvi, \cvi\ and \cv\ 
X--ray lines may be present, the quality of our data is poor at those
longer wavelengths. We therefore do not report measurements for these lines.
The \rgs\ spectra near the \neix\ line is shown in Fig.~\ref{fig:zoom}
and the spectra near the other lines in Fig.~\ref{fig:fit-all}.

\begin{table}
\centering
\caption{Atomic parameters of K$\alpha$ absorption lines (from \citealt{verner1996})
and logarithms of Solar elemental abundances (from \citealt{anders1989},
relative to hydrogen).
 \label{tab:ions}}
\begin{tabular}{llrcc}
\hline
\hline
Ion  & \multicolumn{2}{c}{Wavelength (\AA)} & Osc. strength & Log Solar \\
    &  Rest    & $z=0.0231$ & & Abundance  \\  
\hline 
\nex & 12.134 & 12.41 & 0.416 & -3.91 \\
\neix & 13.447 & 13.76 & 0.724&       \\
\oviii & 18.969 & 19.41& 0.416& -3.07 \\
\ovii & 21.602 & 22.19& 0.696 &       \\
\nvii & 24.781 & 25.35& 0.416 & -3.95 \\
\nvi & 28.787 & 29.45 & 0.675 &       \\
\cvi & 33.736 & 34.52 & 0.416 & -3.44  \\
\cv & 40.267  & 41.20 & 0.648 &  \\
\hline
\hline
\end{tabular}
\end{table}

As noted in Section~\ref{sec:los}, \cite{takei2007} analyzed a sub--set of these observations.
 Their method of analysis differs from ours in a number of ways.
First, a number of updates to the calibration of the data have occurred since the
\cite{takei2007} paper was published. In this paper we apply the available
\xmm\ calibration as of 2016 through the use of the \emph{SAS} software and 
current calibration files (\emph{CCF}).
Another critical point in the analysis is the exclusions of periods of high background.
%as illustrated in Fig.~\ref{fig:flare}. 
Our analysis results in a substantially
shorter clean exposure time, despite the availability of additional observations.
%It is therefore possible that the \cite{takei2007} analysis was in part affected
%by the presence of periods of high background.
Also, they assess the presence of absorption
using the apparent deficit of
photons from a fit to the continuum without lines, using a so--called
`ratio method', whereas
we use a more physically--motivated line fitting method.

\section{Results of the \rgs\ spectral analysis}
In this section we first describe the measurements of the line parameters, then
provide an assessment of the robustness of these measurements with respect to
certain aspects of the analysis.
\label{sec:results}

\subsection{Detection and upper limits of X--ray absorption lines}
\label{sec:detections}
The spectral analysis reported in Table~\ref{tab:spex} 
shows that the \rgs\ spectra towards \xc\ 
are consistent with the presence of a redshifted \neix\ absorption line,
as reported by \cite{takei2007}. 
The feature remains significant at the 90\% confidence level ($\Delta C=2.7$)
with a column density of $\log N (\text{cm}^{-2})= 19.39\pm^{0.57}_{2.39}$ but it is not
significant at the 99\% level ($\Delta C=6.3$), with an upper limit of
$\log N (\text{cm}^{-2}) \leq 20.16$.
Given the limited 
 significance of detection,
we cannot  prove conclusively
the presence of \neix\ absorption lines from the WHIM at the redshift of \coma.
In Section~\ref{sec:WHIM} we further discuss whether a putative \neix\ with the
parameters of Table~\ref{tab:spex} may be
due to WHIM absorption, based on a joint analysis with the
emission detected by \rosat\ \citep{bonamente2003}. 
At the wavelengths of \neix,
RGS2 is the only detector with effective area in our observations. Follow--up observations
with both \xmm\ and  \chandra\ are required to
establish conclusively the reality of this absorption feature, and that of
any additional X--ray absorption lines.

We therefore focus on the upper limits of
WHIM absorption lines allowed by our analysis of the \xmm\ observations.
The upper limits to column densities and optical depths (Table~\ref{tab:spex})
indicate that our sensitivity limits are consistent with
lines that may in fact be saturated ($\tau_0 \geq 1$). In the calculation
of these limits, we have assumed a purely thermal broadening of the lines
with $b=50-100$~km~s$^{-1}$, with \emph{SPEX} taking into account all the atomic physics required to
determine the appropriate line profile, including the effects
of RMS speeds on the line profile (i.e., the presence of saturation).
The WHIM may become turbulent due to
accretion shocks that occur during the infall into the deep
gravitational well of clusters. As a result, 
the WHIM may feature substantially higher non--thermal
speeds up to $\sim$ 1,000~km~s$^{-1}$ \citep[e.g.][]{schmidt2016}.
If the plasma has RMS speeds that are significantly higher than
the thermal speeds assumed in our measurements of Table~\ref{tab:spex},
the lines would become less saturated, leading to \emph{lower}
upper limits to their column density, as also shown for 
the extragalactic source \pg\ in \cite{bonamente2016}.
The upper limits of Table~\ref{tab:spex}
are therefore strict (i.e., the highest) upper limits allowed 
by the available X--ray data. In Sect.~\ref{sec:systematics}
we discuss further the sensitivity of our measurements
to the value of the assumed $b$ parameter.

\begin{table}
\centering
\caption{Results of SPEX fit to high--temperature lines with the \texttt{slab} model.
Confidence levels are 68\%, obtained for $\Delta C = 1$.}
\label{tab:spex}
\begin{tabular}{lllccc}
\hline
\hline
Ion 	& Band (\AA) & $C$ stat (d.o.f) &  \multicolumn{3}{c}{SPEX Fit Results}\\
	&            &			&  $W_{\lambda}$ (m\AA) & $\log N( \text{cm$^{-2}$})$ & $\tau_0$ \\
\hline
\nex 	& 12.0-13.0  & 56.5(47) &$\leq$ 17.0 	& $\leq 16.68$	& $\leq 2.6$\\
\neix 	& 13.2-14.4  & 95.0(57)	& $74\pm^{36}_{31}$ & $19.39\pm^{0.38}_{0.79}$&
				$2.5\pm^{3.1}_{2.6}\times 10^{3}$ \\
\oviii 	& 19.0-20.0  &  116.6(98)	& $\leq$ 29.1 	& $\leq 16.55$	& $\leq 3.0$ \\
\ovii 	& 21.7-22.5  &  32.2(37)	& $\leq$ 16.1 	& $\leq 15.93$ 	& $\leq 1.4$\\
\nvii	& 25.0-25.7  & 89.0(69) 	& $\leq$ 9.2	& $\leq 15.65$ 	& $\leq 0.5$ \\
\hline
\hline 
\end{tabular}
\end{table}

\subsection{Systematics}
\label{sec:systematics}

One of the sources of systematic errors in the analysis of grating spectra
is the uncertainty in the background level for the spectra.
%, the use of
%a fixed width for the Gaussian line model 
%in the reduction and analysis of the data.
%In \cite{bonamente2016} we have shown that the choice of width parameter
%for the Gaussian line model is effectively irrelevant, provided that it is 
%smaller than the resolution of the instrument
%and the absorption line remains on the linear part of the curve of growth. 
%We have used a width parameter
%$\sigma_K=0.2$~eV, corresponding to a broadening of the line
%by $b=100$~km~s$^{-1}$. The width is much smaller than the 
%resolution of the RGS detectors ($\sim 50$ m\AA), therefore
%the results of this paper are insensitive to this choice.
\begin{figure}
\centering
\includegraphics[width=5in]{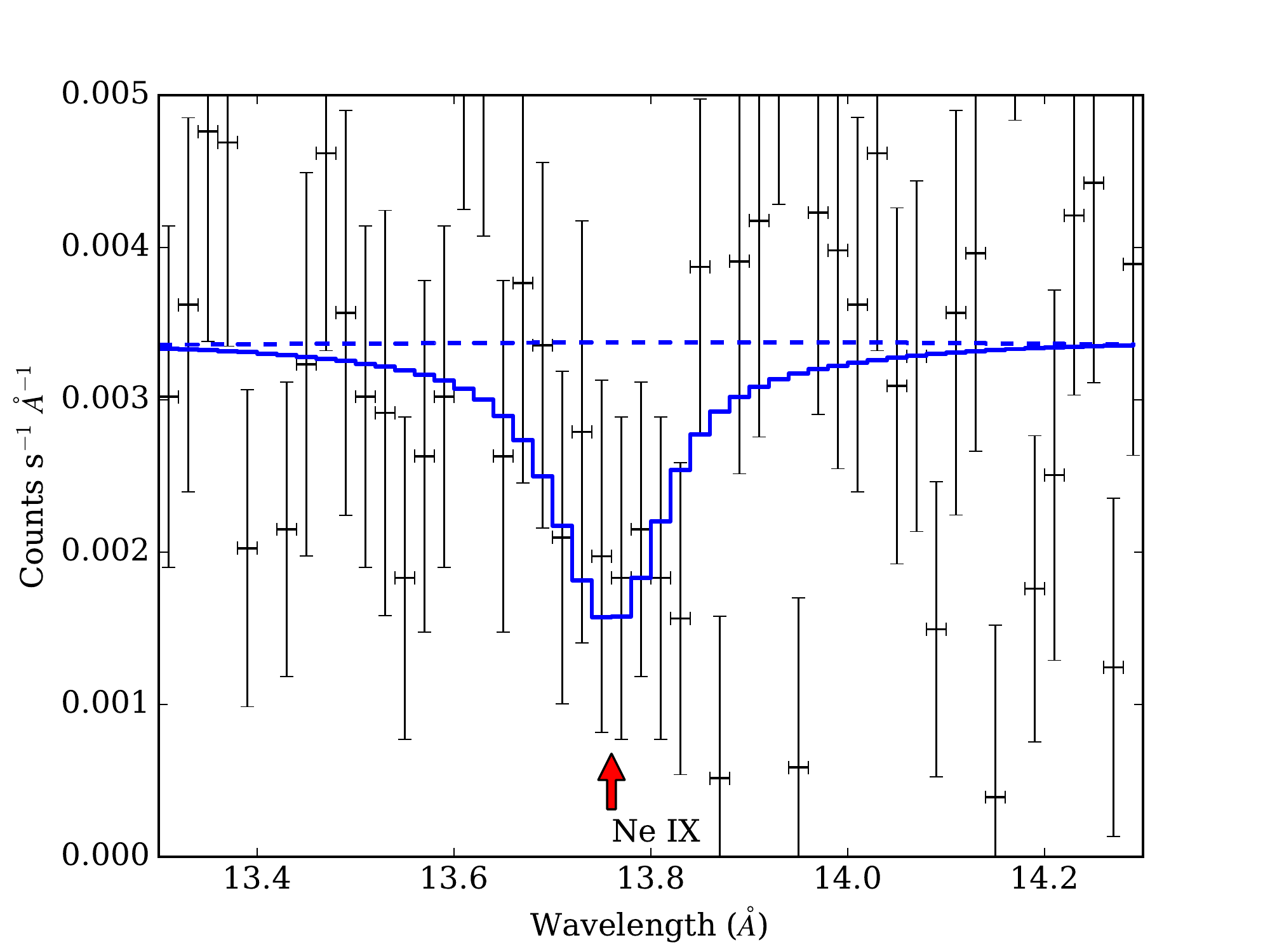}
\caption{First--order RGS2 spectrum of \xc\ at the wavelengths
of the \neix\ K-$\alpha$ line. Dashed line is
the continuum and solid line is the best--fit model with a redshifted \neix\ absorption line.
Spectrum is shown at the resolution of 20 m\AA\ per bin, same as 
in the fit to the spectra.
%in red is the model at the best--fit redshift of  $z_{\neix}=0.0254\pm^{0.0017}_{0.0027}$.
\label{fig:zoom}}
\end{figure}
\begin{figure}
\includegraphics[width=3.5in]{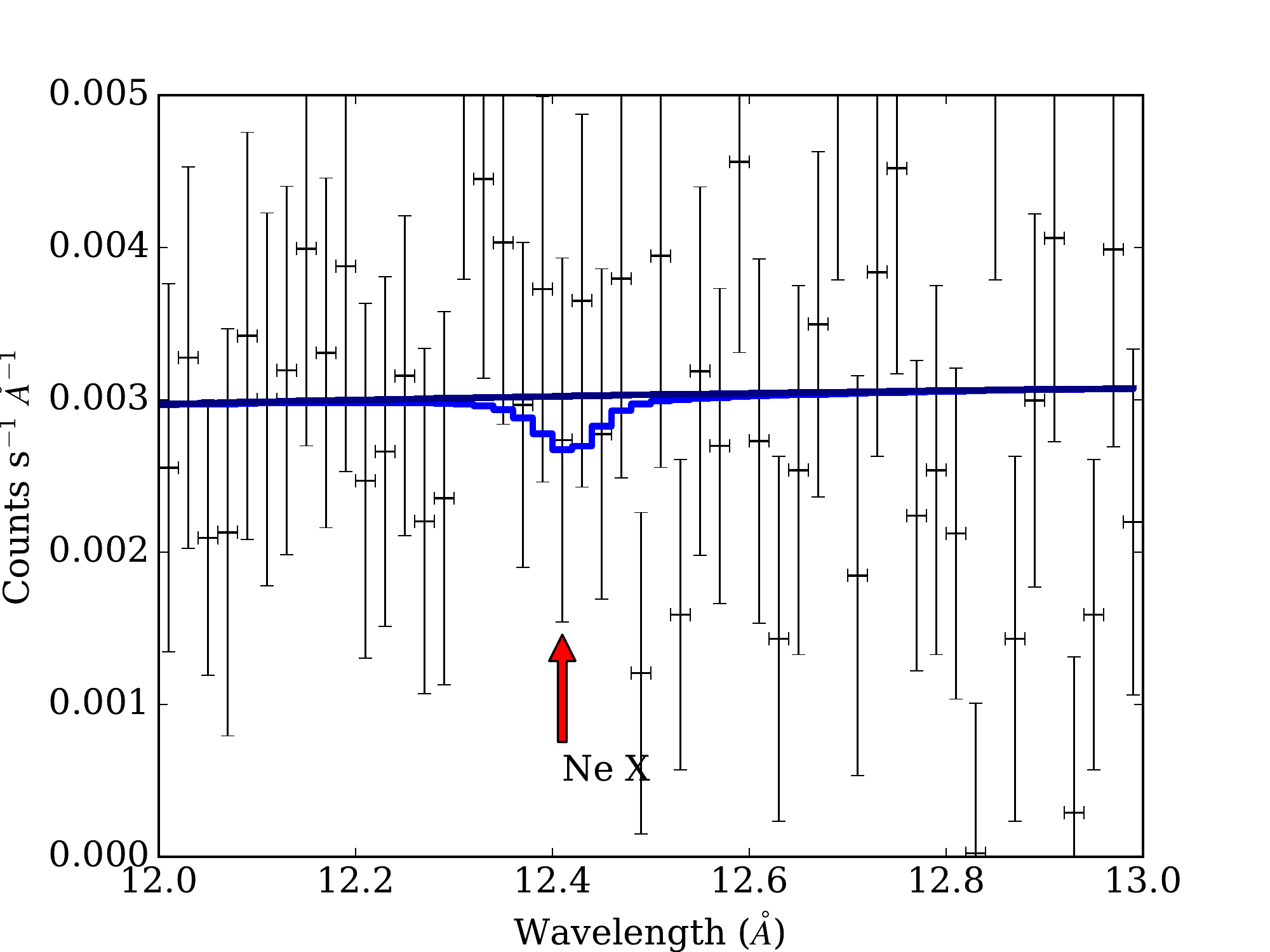}
\includegraphics[width=3.5in]{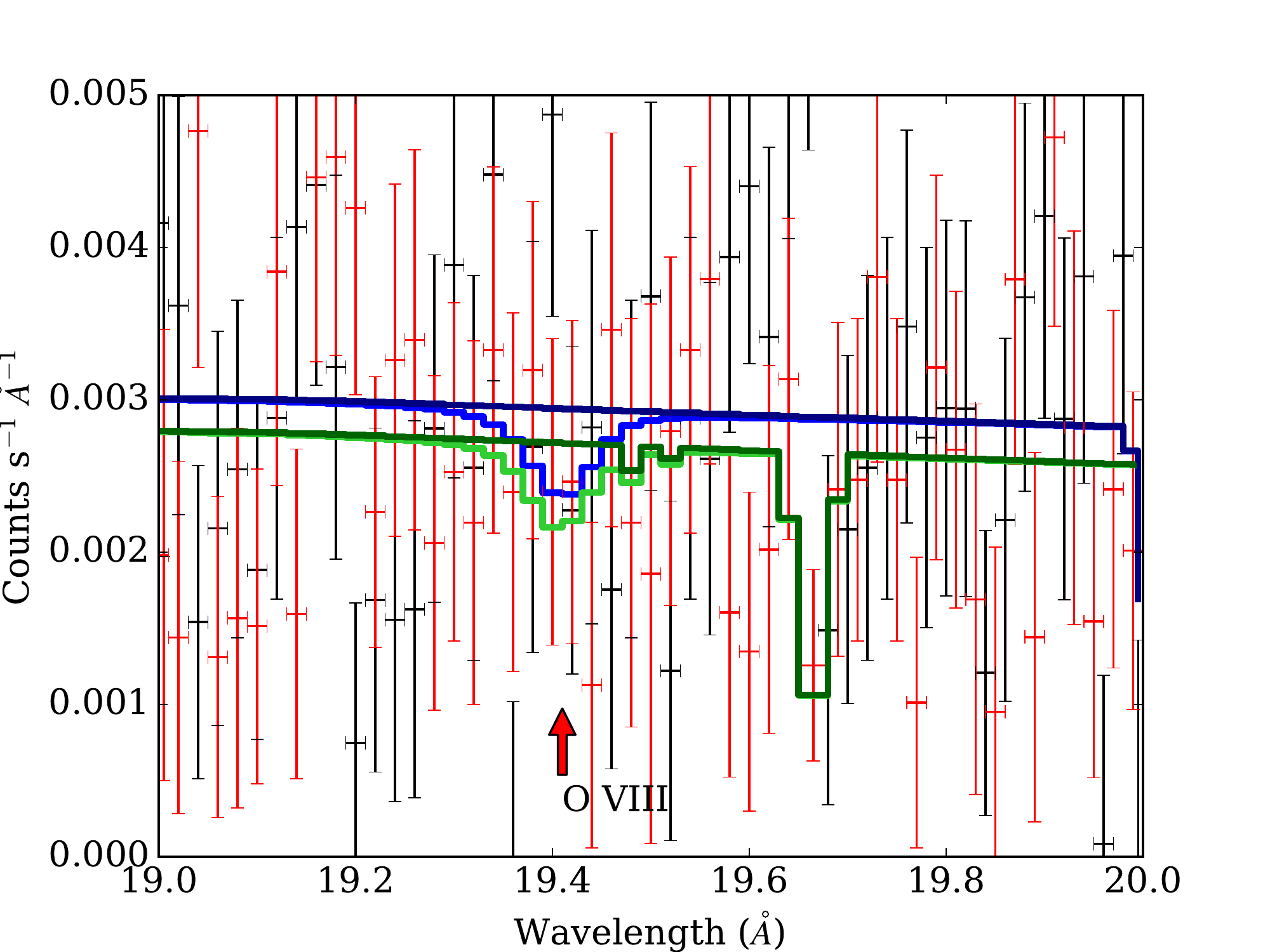}
\includegraphics[width=3.5in]{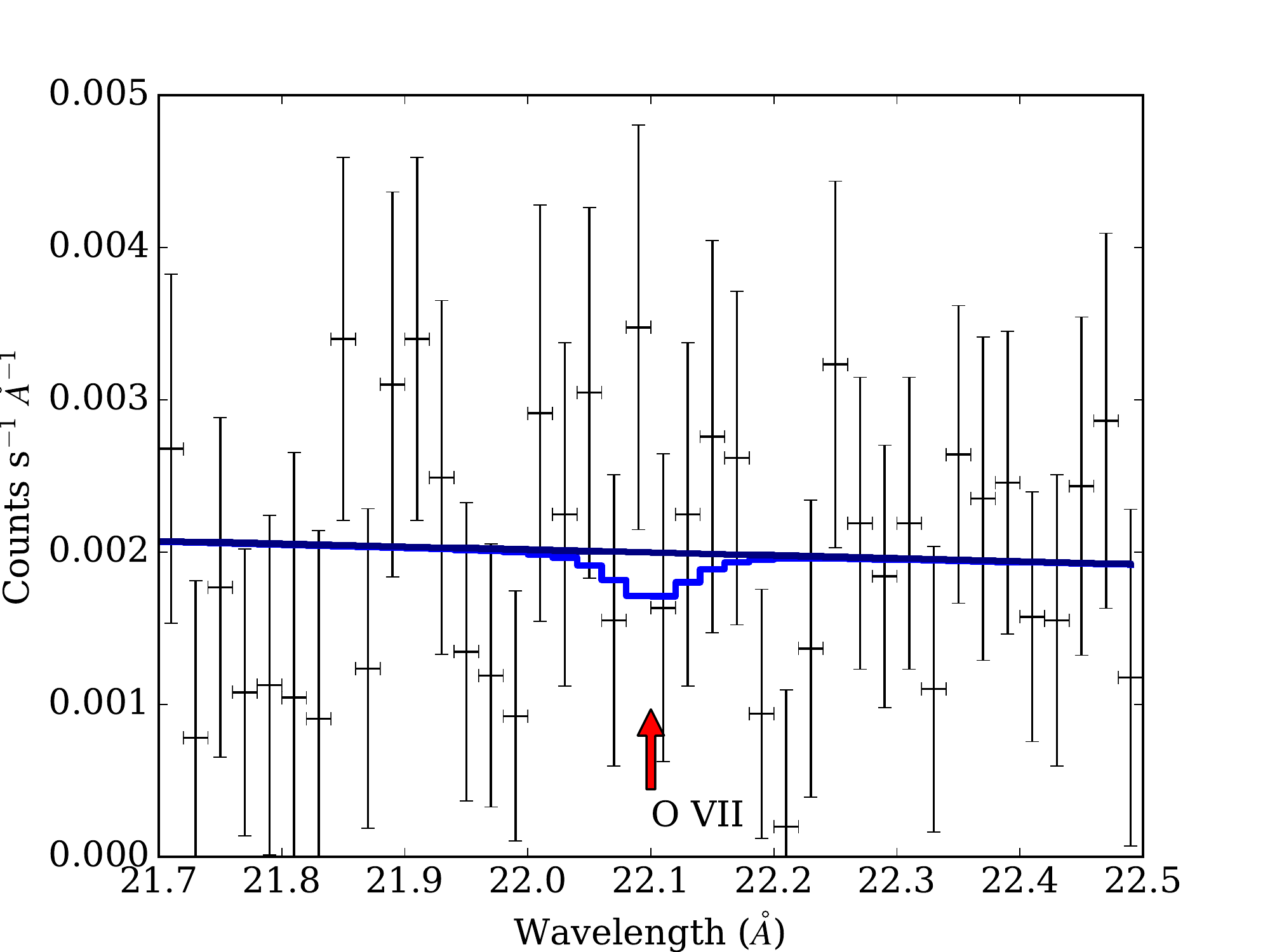}
\includegraphics[width=3.5in]{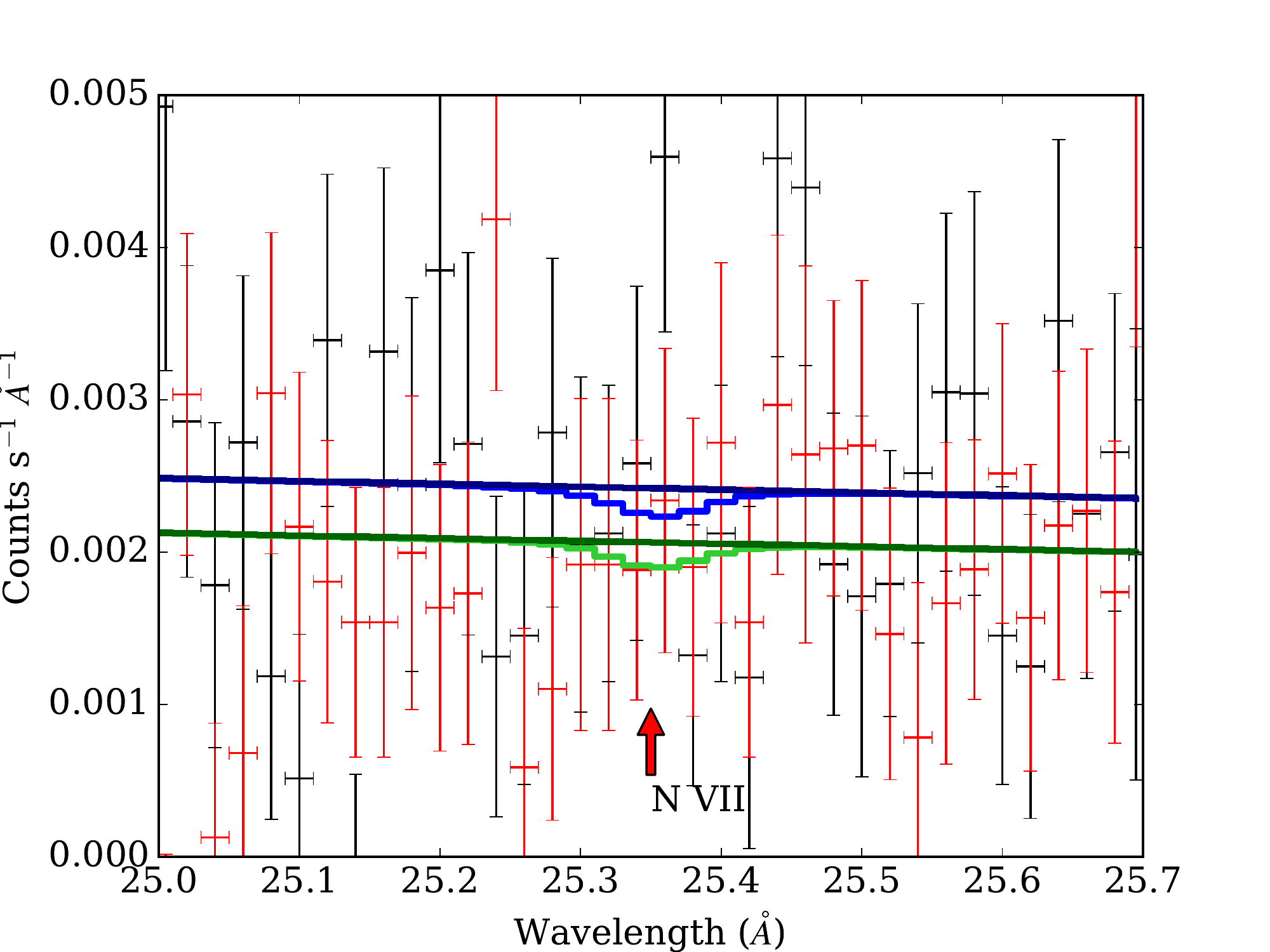}
\caption{First--order spectra of \xc\ and location of 
the  \nex, \oviii, \ovii\ and \nvii\ X--ray absorption lines
at the redshift of \coma\ (z=0.0231).
Dips in the models that do not correspond to labeled features
are instrumental artifacts due to a reduced efficiency of the RGS cameras.
There is no evidence for redshifted absorption from these ions
in the spectra of \xc. Lighter color lines (light green and light blue) correspond to the 68\% confidence 
upper limits of the line column densities allowed by the data.
\label{fig:fit-all}}
\end{figure}
To address systematic errors associated with the background level,
we performed the test of changing the background level by $\pm 10$\%,
as in \cite{bonamente2016}. Near the wavelength of the \neix\ line,
the background is approximately 50\% of the source count rate.
For the main absorption feature (\neix),
we find that the best--fit parameters of the line are effectively
unchanged when the background is changed by this amount, same
as we found in \cite{bonamente2016}.
We therefore conclude that the results of Table~\ref{tab:spex}, including
the upper--limits to non detection, are insensitive to reasonable
variations of background levels.

Another key choice in the reduction of \rgs\ data is the level
of the quiescent background.
%, as shown in Figure~\ref{fig:flare}.
Our choice for the quiescent level (see Section~\ref{sec:data}) was somewhat conservative,
leading to substantially shorter exposure times (for the
same observations) than in the reduction of \cite{takei2007}.
We re--reduced the data using a higher value for the quiescent
background, $0.1$~counts~s$^{-1}$, leading to a new
clean exposure time of 325.1~ks (versus the 187.4~ks of the
standard reduction used throughout the paper). With these new spectra,
the best--fit column density for the \neix\ line becomes
$\log N(\text{cm}^{-2})=19.17\pm^{0.37}_{1.45}$,  consistent 
 with the value of Table~\ref{tab:spex}. For the same high--background data,
we also confirmed that the line is not significant at the 99\% confidence level,
same as for the results with the low--background reduction. We therefore
conclude that our results are robust with
regards to changes in the level of the quiescent background.

We also tested the robustness of the significance of detection of the
\neix\ line with respect to the value of the redshift assumed.
%spectral bin size and redshift.
%For this purpose, we first rebinned the spectra to a bin size of 30~m\AA, and repeated
%the fitting procedure. The \neix\ line remains detected at the same confidence level
%($K=-3.2\pm1.5$, or 2.1 $\sigma$), while the normalization of the
%\oviii\ line  becomes $K=-0.3\pm1.2$, i.e., the significance of this lines falls below the
%$1\sigma$ threshold. We caution the reader that such smaller bin size may render
%some of the bins correlated to one another, and the $\mathcal{C}$ statistic
%may not be accurate to determine the best--fit parameters.
In our analysis we fixed the central wavelength of the lines, according to the
redshift of the \coma\ cluster. It is however possible that infall
or peculiar velocities of WHIM filaments will result in small deviations
from these nominal wavelengths. For this purpose, we re--analyze the spectra
at the wavelengths of the \neix\ line
allowing the redshift to be a free
parameter. We obtain a best--fit redshift of $z_{\neix}=0.0252\pm{0.0020}$,
which is consistent with the nominal redshift of \coma\ ($z=0.0231$)
used throughout this paper.
For this redshift value, the fit statistic is reduced to $C=93.9$ for one
fewer degree
of freedom, and the column density is measured at $\log N(\text{cm}^{-2}) = 19.43\pm^{0.36}_{0.70}$.
Using this redshift, the feature becomes just 
significant even at the 99\% confidence
level ($\Delta C = 6.3$ for one interesting parameter), with a column density
constrained to $\log N(\text{cm}^{-2}) = 19.43\pm^{0.74}_{3.89}$. 
In Fig.~\ref{fig:zoom} we show a close--up of the RGS2 spectrum near the wavelengths
of the putative \neix\ absorption line.

\begin{figure}[!ht]
\centering
\includegraphics[width=5in]{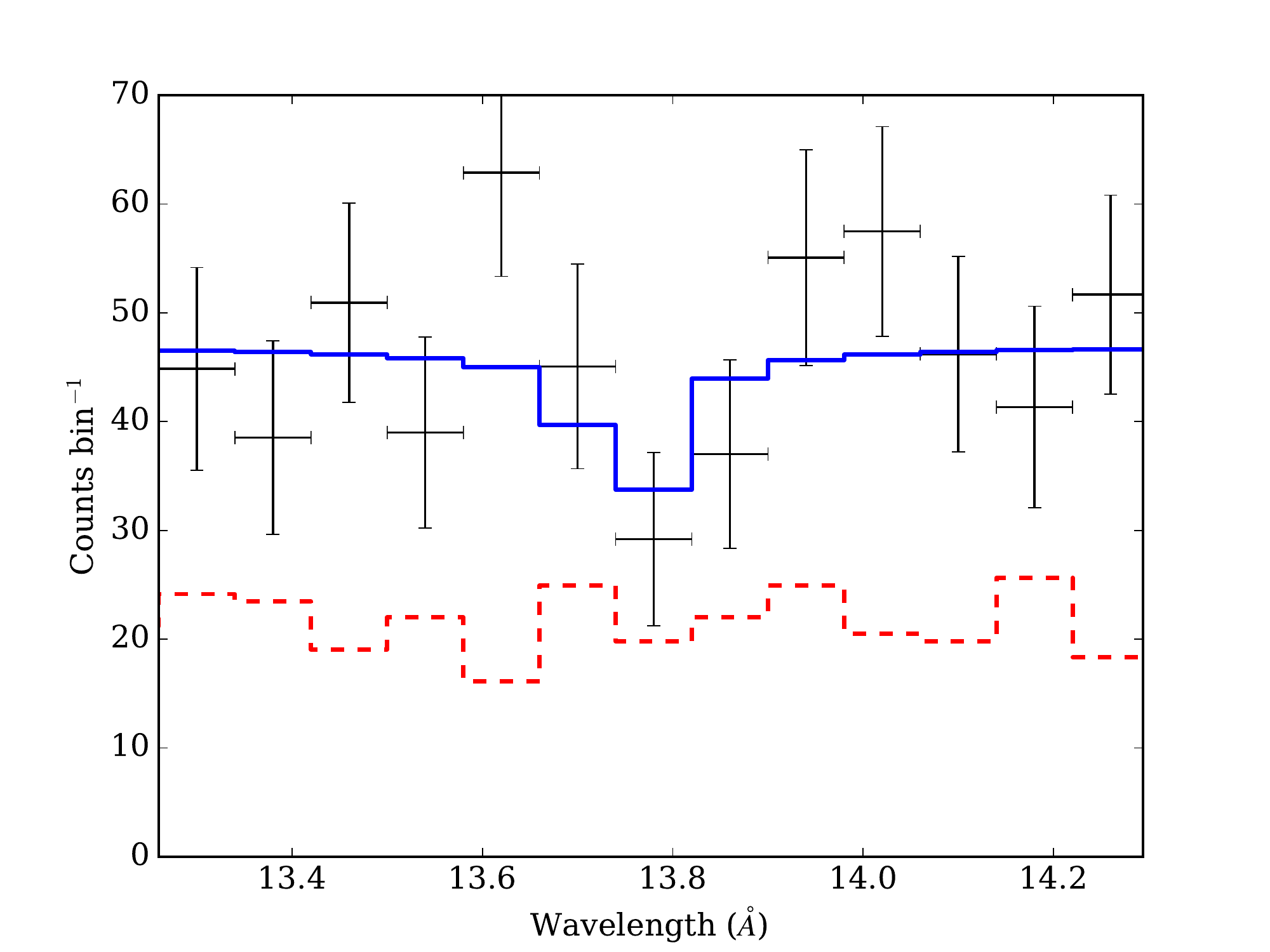}
\caption{First--order \rgs\ spectrum of \xc\ rebinned by a factor of 8, to ensure
that there are at least 20 counts in each bin (notice the units on the $y$ axis).
The best--fit model in blue is obtained by a fit with the $\chi^2$ statistic.
In red is the subtracted background.}
\label{fig:chi2}
\end{figure}

The significance of detection of the putative \neix\ line was tested also
by rebinning the spectrum by a factor of 8 (or 80~m\AA), so that each bin has at least
20 counts, and using the $\chi^2$ statistic in the fit.  With this coarser
binning, the spectrum in the vicinity of the redshifted line (see Figure~\ref{fig:chi2})
yields a best--fit statistic of $\chi^2_{\text{min}}=9.5$ for 11 degrees of freedom, and 
the best--fit \neix\ column density is $\log N (\text{cm}^{-2})= 19.19\pm^{0.53}_{1.66}$ 
(68\% error), consistent with the results obtained with the Cash statistic
and the spectra binned by a factor of two. Even with this coarser binning, the feature
remains consistent at the 90\% level (or $\Delta \chi^2=2.7$ for one degree of freedom),
with a confidence interval of $\log N (\text{cm}^{-2})= 19.19\pm^{0.77}_{3.16}$.
The feature is not significant at the 99\% level, also consistent with the Cash statistic analysis.

Finally, we address the sensitivity of our measurements to
the choice of the velocity $b$ parameter, which we have set to 
a value $b=50-100$~km~s$^{-1}$, corresponding to purely
thermal broadening of the lines. As discussed in Sect.~\ref{sec:detections},
a larger value for $b$ is in principle possible if the WHIM has a significant
non--thermal velocity structure.
To discuss the effect of $b$ on our measurements, we 
fit the spectra in the neighborhood of the \neix\ line with
a fiducial value of $b=500$~km~s$^{-1}$. The goodness of fit is unchanged
relative to the fit with thermal broadening only, since \rgs\ does not
have sufficient spectral resolution to detect this level of line broadening.
As a result, the line's equivalent width is nearly 
unchanged ($W_{\lambda}=70\pm^{26}_{27}$ m\AA),
but the best--fit column density and optical depth are 
both significantly reduced, respectively  to a value of $\log N(\text{cm}^{-2})=
17.08\pm^{0.42}_{0.36}$
and $\tau_0=2.5\pm^{4.1}_{1.4}$. This is due to the fact that 
a larger velocity structure keeps the line on the linear portion of
its curve of growth for larger values of the column density. Likewise,
upper limits to other lines would be correspondingly reduced by the
assumption of a larger $b$ parameter. For example,
a value of $b=500$~km~s$^{-1}$ results in 
an upper limit to  the \ovii\ line of $\log N(\text{cm}^{-2}) \leq 15.82$,
corresponding to $\tau_0 \leq 0.21$ and $W_{\lambda} \leq 17.7$~m\AA.
This reduction in the column density of \ovii, if significant non--thermal
velocities are present, is more modest than that for \neix. In fact,
the values in Table~\ref{tab:spex} indicated that even for a purely thermal
velocity, there is only mild saturation of \ovii\ K$\alpha$ lines.

\section{Interpretation}

In this section we discuss the findings of our analysis  
of the \rgs\ spectra of \xc. First we analyze the distribution of galaxies
along the sightline, and conclude that there is no individual
galactic halo with a column density that is consistent with our
measured upper limits. We then provide a joint
interpretation for the upper limits to the
non--detection of X--ray absorption lines and the X--ray emission
detected by \rosat\ to constrain the physical parameters of putative
WHIM near the \coma\ cluster. We also discuss whether a \neix\ absorption
line with the characteristics of Table~\ref{tab:spex} (i.e.,
highly saturated and with large column density) is consistent with a WHIM origin.

\subsection{Galaxy Confusion}

\begin{figure}[!t]
\centering
\includegraphics[width=6in]{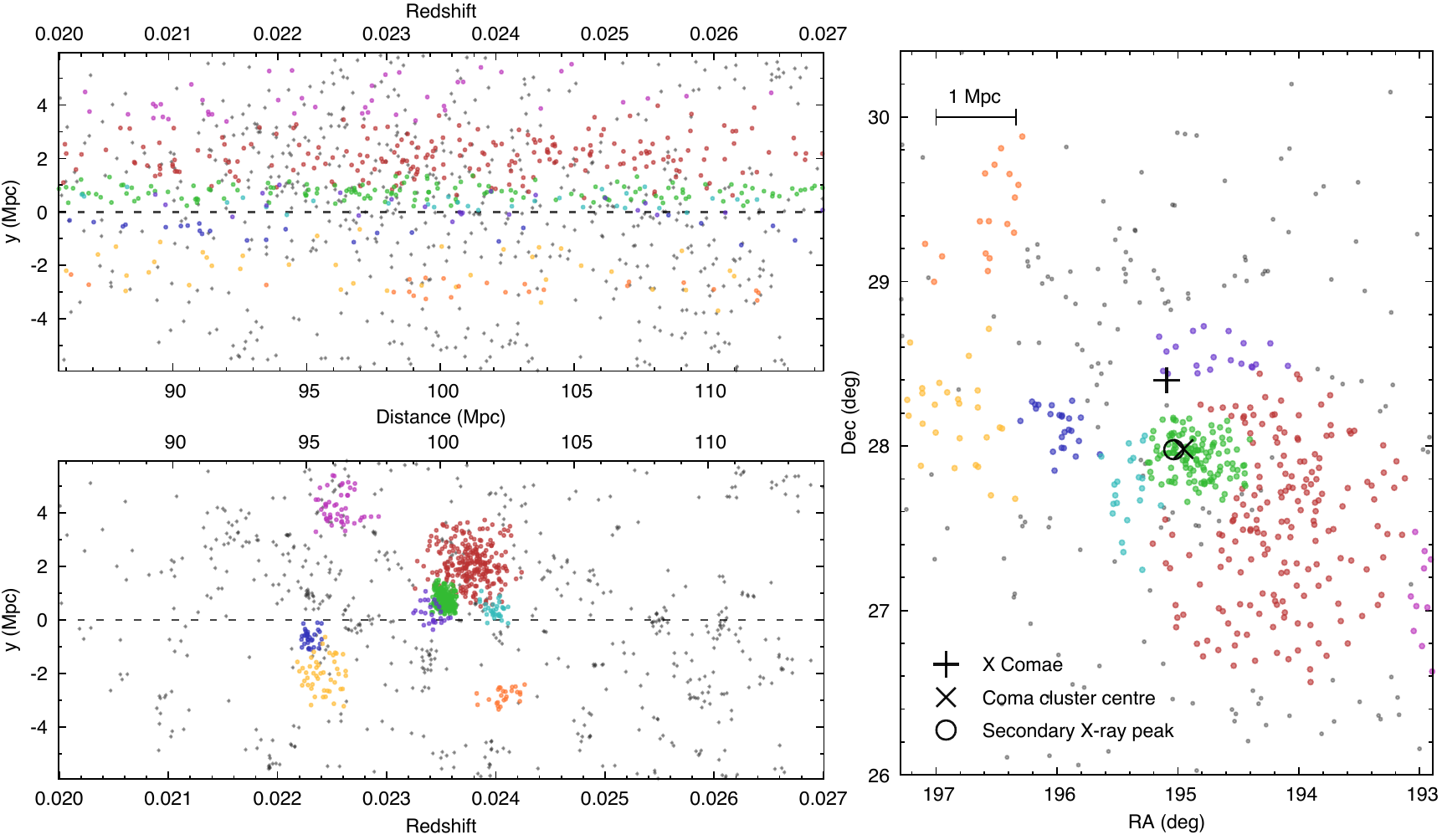}
\caption{Distribution of galaxies around \xc\ in
the \emph{SDSS} survey. Right panel shows the distribution of galaxies
(redshifts between 0.02 and 0.027) in the sky plane.
\xc's  line of sight is marked as a cross and the secondary X-ray peak is marked with a circle.
Upper left panel shows the distribution of galaxies along the \xc\ line of sight.
Bottom left panel shows the distribution of galaxies after suppression of the Fingers-of-God effect
(see text for more details). Each galaxy system with at least 20 members are shown with
various colours, all other galaxies are shown as grey points.
}
\label{fig:sdss}
\end{figure}

We first study the distribution of galaxies from optical \sdss\ data to investigate galaxy confusion along
the sightline towards \xc.
\coma\ is a massive cluster ($M \geq 10^{15} M_{\odot}$, \citealt{geller1999})
with hundreds of galaxies within a radius of a few Mpc.
Fig.~\ref{fig:sdss} shows the distribution of galaxies around the \xc\ sightline.
Galaxies are selected from the \sdss\ main spectroscopic sample with apparent magnitudes
brighter than 17.77 in the $r$--band.
To suppress the Fingers-of-God effect, a redshift--space distortion in which the observed galaxy
distribution appears elongated in redshift space along the line of sight \citep{jackson1972,tully1978},
we applied the modified Friends--of--Friends algorithm \citep{tempel2016}
and detected systems of galaxies with at least two members. We then calculated the velocity dispersion
along the line of sight and the group extent on the sky plane. These quantities were used to
spherize the galaxy groups while altering the distances of galaxies (see  \citealt{tempel2014b} for more details).
Left panels in Fig.~\ref{fig:sdss} illustrate the Fingers--of--God correction.
The sightline towards \xc\ therefore interesects a number of galaxies and groups that are members of
the \coma\ cluster.

In \cite{nevalainen2015} we showed that a galaxy of $r$--band luminosity similar to
the Milky Way ($m_r \simeq -22$) requires an impact parameter of $\leq$ 10 kpc to produce
a column density of $\log N_H (\text{cm}^{-2}) = 19$ within the sightline. Our analysis of the \sdss\ data
show that there are no galaxies closer than 70 kpc to the \xc\ sight line at the redshift of
the absorber. Thus, spirals in the sample are not dense and close enough to produce
the measured upper limit of $\log N_H (\text{cm}^{-2}) \sim 21$ 
(see Section~\ref{sec:results}).
There are four bright elliptical galaxies with $m_r \leq -22$ along the sightline.
For these, we used \cite{kim2013}, Fig.~2, to estimate their X--ray luminosity in the range
 $\log L_X (\text{erg s$^{-1}$})= 40-41$, based on the $r$--band luminosity.
We then use  \cite{fukazawa2006}, Fig. 3, to estimate their gas density from the X--ray luminosity,
which we find to be in the range $\log n_H (\text{cm$^{-3}$})= (-4,-5)$, at the impact parameters
of these galaxies. Such densities would require a pathlength in excess of 1~Mpc, i.e.,
larger than the virial radii of the galaxies, to produce a
column density of the order of $\log N_H (\text{cm}^{-2}) = 21$. We conclude that there is no suitable elliptical
galaxy that can provide a comparable column density to the measured upper limits.

\subsection{Constraints on the presence of WHIM filaments near the \coma\ cluster}
\label{sec:WHIM}
The sightline towards \xc\ is rendered particularly interesting
by the presence of an excess of soft X--ray radiation around the 
\coma\ cluster. In \cite{bonamente2003} we have found evidence for
the presence of $\sim \nicefrac{1}{4}$~keV plasma that extends
well beyond the projected radius of the \xc\ sightline
\citep[see also][]{bonamente2009}. 
This soft X--ray emission detected by \rosat\  can be interpreted with the presence
of WHIM filaments converging towards \coma.
In this section
we combine the upper limits to the X--ray absorption lines with
the positive detection of emission near \coma\ we obtained with \rosat\
to derive properties
of the putative WHIM near the \coma\ cluster. 

\subsubsection{Constraints from \rosat\ emission and \rgs\ absorption lines upper limits}
 Assuming
a filament of uniform density $n_H$ and length $L$ along the sightline,
its soft X--ray emission integral is given by
\begin{equation}
\int n_e n_H dV = I \times \left(10^{12} 4 \pi D_A^2 (1+z)^2\right),
\label{eq:emission}
\end{equation}
where  the right--hand side term in parenthesis is referred to
as the constant $C_I$, and the emission integral $I$ was measured
in \cite{bonamente2003} for a quadrant between radii 20--40',
corresponding to the projected radial distance of \xc\ from the
cluster's center. The measurement was $I=0.7\pm0.08$ in $cgs$ units 
(see Table~3 of \citealt{bonamente2003}).
For our simple uniform--density filament model
this equation becomes

\begin{equation}
n_H^2 \left(\frac{\mu_H}{\mu_e}\right) L S = I C_I
\label{eq:emission2}
\end{equation}
where $S$ is the surface area of the filament (so that $L S$ is the volume),
$\mu_H/\mu_e=n_e/n_H \simeq 1.2$.
The redshift $z=0.0231$ and the angular distance $D_A$
are those of the \coma\ cluster. The surface area of the
filament, for the chosen emission region, is $S=\nicefrac{1}{4}\; \pi D_A^2 (\theta_2^2-\theta_1^2)$,
where the angles of respectively 40' and 20' are measured in radians, and the
factor of $\nicefrac{1}{4}$  accounts for the shape of the region 
(a quadrant of an annulus).
Equation~\ref{eq:emission2} thus links the  density and length of the putative WHIM filament.

From upper limits to the presence of absorption lines we can
further set
\begin{equation}
N_H = n_H L \leq N_{H,UL}(T)
\label{eq:absorption}
\end{equation}
where the temperature--dependent 
upper limits to the hydrogen column density $N_{H,UL}(T)$
can be obtained from the measured upper limits for the various
ions of Table~\ref{tab:spex}. Such upper limits are 
obtained by assuming a chemical abundance (e.g.,
$A=0.1$ Solar) and the temperature--dependent
ionization fractions $f_{\text{ion}}(T)$ calculated in ionization equilibrium, via
\[
 N_{H,UL}(T) = \frac{N_{\text{ion},UL}}{f_{\text{ion}}(T) A}.  
\]
The ionization fractions are very sensitive to the choice
of temperature. To account for uncertainties in the
 WHIM temperature towards \xc,
we use a temperature of $kT=0.22\pm0.05$~keV ($T=2.5\pm0.6 \times 10^{6}$~K), 
corresponding
to the best--fit value 
and a conservative 3--$\sigma$ error bar
from the \rosat\ analysis of \cite{bonamente2003}.
We calculate the ionization fractions of \nex, \neix,
\oviii, \ovii\ and \nvii\  assuming collisional ionization equilibrium at this
temperature \citep{gnat2007}. We report the ionization fractions and constraints to the hydrogen
column density  in Table~\ref{tab:colDens}, mindful of the fact that these
limits are inversely proportional to the value of the chemical abundance
(assumed 0.1 Solar).

\begin{table}[!t]
\centering
\caption{Limits to column densities of WHIM filaments towards \xc \label{tab:colDens}.
Abundances are assumed to be 10\% Solar, using the \cite{anders1989} Solar values
from Table~\ref{tab:ions}. Ionization fractions in CIE are for an assumed temperature
of $kT=0.22\pm0.05$~keV \citep{bonamente2003}.\label{tab:filaments}}
\begin{tabular}{lccl|lll}
\hline
    &                &                  &          & \multicolumn{3}{c}{Constraints on WHIM Filament Properties} \\
Ion & $f_{\text{ion}}$ & $\log A$ & $\log N_{H,UL}$  	& Length $L$  & \multicolumn{2}{c}{$H$ Density}\\
    &           &          & (cm$^{-2}$)  	&  (Mpc) 	& $n_H$ (cm$^{-3}$) & $\delta_b$ \\    
\hline  
\nex   & $0.07\pm^{0.09}_{0.06}$ & -4.91 &  $\leq 22.7\pm^{0.9}_{0.4}$ & 
	\multicolumn{3}{c}{(Uninteresting limits)}\\
\neix  & $0.93\pm^{0.05}_{0.13}$ & -4.91 & $24.3\pm^{0.4}_{0.8}$ & $(0.1-22)\times10^8$ & 
	$0.8-11 \times 10^{-9}$ &$0.003-0.04^{\star}$\\
\oviii & $0.45\pm^{0.00}_{0.12}$ & -4.07 & $\leq 21.0\pm0.1$	&$\leq 71\pm^{62}_{0}$ & 
	$\geq 4.1\pm^{0.0}_{1.1} \times 10^{-6}$ &$ \geq 14.4\pm^{0.0}_{3.9}$ \\
\ovii  & $0.28\pm^{0.35}_{0.18}$ & -4.07 & $\leq 20.6\pm^{0.5}_{0.4}$ & $\leq 11\pm^{73}_{9}$ & 
	$\geq 1.0\pm^{1.2}_{0.7} \times 10^{-5}$ &$ \geq 37\pm^{47}_{24}$ \\
\nvii  & $0.21\pm^{0.22}_{0.11}$ & -4.95 & $\leq 21.3\pm0.3$ & 
	\multicolumn{3}{c}{(Uninteresting limits)}\\
\hline
\end{tabular}
\flushleft
\footnotesize{$\star$ Constraints from \neix\ assume that the redshifted \neix\ K$\alpha$ detection is real}.
\end{table}

Using Eq.~\ref{eq:emission2} into Eq.~\ref{eq:absorption}, the combination
of soft X--ray emission from the WHIM and upper limits to absorption lines leads to
\emph{upper limits} to the filament's lenght,
\begin{equation}
L \leq N_{H,UL}^2 \frac{S}{I} \left(\frac{\mu_H}{\mu_e}\right) \left(\frac{A}{0.1 \text{ Solar}}\right)^{-2},
\label{eq:L}
\end{equation}
and \emph{lower limits} to the WHIM density,
\begin{equation}
n_H \geq \frac{I}{S} \frac{1}{N_{H,UL}} \left(\frac{\mu_e}{\mu_H}\right) 
\left(\frac{A}{0.1 \text{ Solar}}\right),
\label{eq:density}
\end{equation}
These limits are reported in Table~\ref{tab:filaments}. 
We interpret the density constraints also in terms
of the critical density of matter of the universe, calculated as 
\[ \rho_{crit} = \frac{3 H^2}{8 \pi G}
\]
where $H$ is the Hubble parameter and $G$ is the gravitational constant. For a flat
universe with $H_0=70$~km~s$^{-1}$~Mpc, the critical density
at present epoch corresponds to $\rho_{crit}=5.5\times 10^{-6}$~H atoms~cm$^{-3}$.
For a baryon density of $\Omega_b=0.05$, the baryon overdensity associated
with the WHIM densities measured from Eq.~\ref{eq:density} is $\delta_b=n_H/( \rho_{crit} \Omega_b)$,
also reported in Table~\ref{tab:filaments}.

For lines with upper limits, 
the most stringent constraints on the properties of the WHIM filaments 
are obtained
from the measurements of the \ovii\ and \oviii\ K$\alpha$ lines. These limits
assume a WHIM temperature of $kT = 0.22\pm0.05$~keV (as measured
by \citealt{bonamente2003}) and a chemical abundance of 
0.1 Solar. The results indicate that the WHIM in the neighborhood of \coma\ may be relatively dense
($n_H \geq 10^{-6}$~cm$^{-3}$ or $\delta_b\geq 10$) and spread over a line--of--sight
distance of tens of Mpc. Both constraints are consistent with
the expected properties of the WHIM. Upper limits from \nex\ and \nvii\ provide less stringent constraints
because of the assumed temperature. Implications of the \neix\ measurements 
for the WHIM scenario are discussed below in Sect.~\ref{sec:neix}.

\subsubsection{Is the \neix\ \rgs\ absorption real?}
\label{sec:neix}
In Table~\ref{tab:filaments} we also reported the WHIM filament constrains from the
\neix\ absorption line, based on the assumption of a positive detection of highly--saturated \neix,
as obtained from our SPEX analysis
(Table~\ref{tab:spex}). Such
constraints on filament length and density are obtained by replacing the inequality signs with equal signs in both
Eq.~\ref{eq:L} and \ref{eq:density}.
The constraints on filament length exceed the size of the universe, i.e., clearly a \neix\
absorption line with the parameters of Table~\ref{tab:spex} is untenable.
This analysis therefore excludes the possibility that
there exists highly--saturated \neix\ absorption (at $b=50-100$ km~s$^{-1}$) that originates from
the WHIM medium at $kT=0.22\pm0.05$ 
and that is responsible for the \rosat\ emission.

A simple explanation for this interpretational problem is that the \neix\ is not real, and the
absorption is a statistical fluctuation. It is nonetheless possible that
there is a significant column density of \neix\ along the sightline to \xc, as
suggested by the \cite{takei2007} study of both emission and absorption.
If the putative absorbing plasma has a very large non--thermal velocity,  the
\neix\ column density of Table~\ref{tab:spex} would be significantly reduced,
as the \neix\ K--$\alpha$ line would be less saturated. 
We have analyzed this possibility in Sect.~\ref{sec:systematics}, where
we found that using a fiducial value of $b \sim 500$ km~s$^{-1}$ 
for the broadening of the line,
the column density is reduced to $\log N(\text{cm}^{-2})=
17.08\pm^{0.42}_{0.36}$.
Even with this lower value, however, the required WHIM filament length 
(calculated in Table~\ref{tab:filaments} for the larger column density) 
remains in excess of 1 Gpc. The WHIM explanation for the 
\rgs/\rosat\ data appears not tenable for any reasonable
value of the line broadening paramater.

\subsubsection{Consistency between the \rgs\ \neix\ absorption and the 
	\emph{EPIC} \neix\ emission of \cite{takei2007}}

In this section we investigate the consistency between the measurements
of the redshifted \rgs\ \neix\ line presented in this paper, with the 
possible detection of \neix\ \emph{emission}
based on the analysis of the lower--resolution \xmm\ \emph{EPIC} data
by \cite{takei2007}. 

If the \cite{takei2007} excess of photons near the \neix\ line is 
interpreted as an emission line at the \coma\ redshit, the measured surface brightness 
$I=2.5\pm1.2 \times 10^{-8}$~photons~cm$^{-2}$~s$^{-1}$~arcmin$^{-2}$ of
the putative \neix\ line is proportional to the average WHIM density $n_H$ and length $L$ according to 
\[ I = \frac{B}{(1+z^3)}Z n_H^2 L \]
where $B$ is a temperature--dependent coefficient 
that equals $B \sim 5 \times 10^{-20}$ photons~cm$^3$~s$^{-1}$~arcmin$^{-2}$ 
for a temperature of $4 \times 10^6$~K \citep[see Eq. 9 in][]{takei2007}.
Assuming 10\% Solar abundances of neon, this converts to the following constraint
between WHIM density and length,
\begin{equation}
 \left(\frac{n_H}{10^{-5}~\text{cm}^{-3}} \right)^2 \left(\frac{L}{10~\text{Mpc}} \right) 
	\sim 10 \left( \frac{A}{0.1 \text{ Solar}} \right)^{-1},
\label{eq:takei}
\end{equation}
i.e., for a density slightly  of $10^{-5}~\text{cm}^{-3}$, the 
emitting region would have a length of
 order $L=100$~Mpc for a neon abundance of 10\% Solar, or
a lenght of $L=10$~Mpc for Solar neon abundance.

We now investigate the consistency between the \cite{takei2007} \neix\ emission line 
(e.g., the constraints from Eq.~\ref{eq:takei}) and our \rgs\ absorption line measurements. 
For this purpose,  the constraints in $n_H-L$ space
from the \rgs\ absorption can be obtained from
\begin{equation}
 n_H L = \frac{N}{f_{\text{ion}} A}
\label{eq:nh-L}
\end{equation}
where $N$ is the column density of the ion of interest (in this case \neix),
$f_{\text{ion}}$ and $A$ respectively the ion fraction and the corresponding 
chemical abundance (i.e., 10\% Solar).
For a value of $\log N (\text{cm}^{-2})= 19.39$, 
%\pm^{0.57}_{2.39}$,
obtained for the case of a purely thermal line broadening, 
the constraints according to Eq.~\ref{eq:nh-L} are
\begin{equation}
 \left( \frac{n_H}{10^{-5}\; \text{cm}^{-3}} \right) \left( \frac{L}{10 \;\text{Mpc}} \right) = 6.7 \times 10^{3} 
	\left( \frac{A}{0.1 \; \text{Solar}} \right)^{-1} f_{\text{ion}}^{-1}.
\label{eq:nh-L-high}
\end{equation}
The combination of Eq.~\ref{eq:takei} (constraints from the emission line) 
and \ref{eq:nh-L-high} (constraints from the saturated absorption line) 
yields a solution for both density and filament
length of
\begin{equation*}
\begin{cases}
  \left( \dfrac{n_H}{10^{-5}\; \text{cm}^{-3}} \right) = 1.4 \times 10^{-3} f_{\text{ion}} \\
 \left( \dfrac{L}{10 \; \text{Mpc}} \right) = 5 \times 10^6  \left( \dfrac{A}{0.1 \;\text{Solar}} \right)^{-1} f_{\text{ion}}^{-2}.
\end{cases}
\end{equation*}
It is clear that for any plausible value of the metallicity and the
\neix\ ion fraction, the solution is not tenable. The ionization 
fraction is in fact $f_{\text{ion}} \leq 1$ at all temperatures, and only
implausibly large abundances of neon would reduce the required filament
length.

We next repeat the same calculations assuming the larger value of the line broadening ($b= 500$~km~s$^{-1}$),
for which we measured an absorbing column of  $\log N(\text{cm}^{-2})=
17.08 \pm^{0.42}_{0.36}$. 
In this case we obtain a constraint from absorption of
\begin{equation}
 \left( \frac{n_H}{10^{-5}\; \text{cm}^{-3}} \right) \left( \frac{L}{10 \;\text{Mpc}} \right) = 32\pm^{50}_{12}
        \left( \frac{A}{0.1 \; \text{Solar}} \right)^{-1} f_{\text{ion}}^{-1}
\label{eq:nh-L-low}
\end{equation}
with a solution for both emission and absorption of
\begin{equation*}
\begin{cases}
 \left( \dfrac{n_H}{10^{-5} \; \text{cm}^{-2}} \right) = 0.3\pm{0.2}\; f_{\text{ion}} \\
 \left( \dfrac{L}{10 \; \text{Mpc}} \right) = 2.3\pm^{3.5}_{1.0} \times 10^4  \left( \dfrac{A}{0.1 \; \text{Solar}} \right)^{-1} 
f_{\text{ion}}^{-2}.
\end{cases}
\end{equation*}
The density of the filaments becomes reasonable, but the 
values of the length remain extreme, unless there is a significantly super--Solar
abundance of neon.

An additional consideration for this WHIM interpretation is provided by the possible
clumpiness of the WHIM. If the WHIM is not homogeneous, as assumed in our calculations,
the emission is boosted by the clumpiness factor 
$C = <{n_H^2}> / <{n_H}>^2$, with $C \geq 1$, 
while the properties of the absorption would remain unaffected.
This results in a factor of $C$ in the left--hand side of Eq.~\ref{eq:takei},
a factor of $C^{-1}$ in the solution for density and a factor
of $C$ in the solution for length. In other words, a clumpy medium
will require smaller densities and even longer filaments
to explain the emission and absorption  measurements. 
Clumpiness will therefore exacerbate the
interpretational problems within the contest of a WHIM origin.
Given the challeges in the intepretation of the emission and absorption
from the WHIM, we conclude that it is unlikely that both
features (\neix\ absorption and \neix\ emission) are real.
\subsubsection{Other considerations}

Studies of a possible filamentary structure containing
WHIM from the distribution of galaxies, e.g., from the \emph{SDSS},
is limited, since our filament-finding algorithm \citep{tempel2014a} is not optimised to work
in complicated cluster environments (e.g., Fig.~\ref{fig:sdss}). Thus, the task of identifying the WHIM absorber
with a large scale galaxy filament from the available \emph{SDSS} data was inconclusive.

A filament with these properties will only contribute
marginally to the \emph{Sunyaev--Zeldovich} effect (SZE) towards \coma, which is
proportional to the integrated pressure along the sightline.
In fact, a 20~Mpc filament with an overdensity of $\delta_b = 50$
and temperature $\nicefrac{1}{4}$~keV
has a Compton-$y$ parameter of just $y \simeq 3 \times 10^{-7}$, approximately
two orders of magnitude smaller than what \planck\ measured
at the projected radius of \xc\  \citep{planck2013-coma}.

\cite{nicastro2016b} pointed out that some redshifted absorption lines from the WHIM may
instead be lines from lower--ionization ions such as \oi\ or \oii\ at zero redshift, i.e., 
located in the Galaxy. Recently, there
has been much interest in the calculation of cross--sections for the 
inner--shell absorption 
of low--ionization atoms of cosmological interest, such as oxygen and neon 
\citep[e.g.][]{gatuzz2013,gatuzz2015,muller2017}. 
The Lyman--series lines from
oxygen ions (\oi\ through \oviii) reported in \cite{nicastro2016} and also in \cite{gatuzz2013} are not relevant to the present
\neix\ K$\alpha$ absorption line,
since they are much lonwgards of this putative redshifted \neix\ line.
The works of \cite{gatuzz2015} and \cite{muller2017} also indicate a \neii\ K$\alpha$ line at $\lambda \simeq 14.6$ \AA\
and \neiii\ K$\alpha$ line at $\lambda \simeq 14.5$ \AA, with a $K$ edge at $\lambda \simeq 14.3$ \AA\
(also reported in \citealt{gatuzz2013}). It therefore appears that low--ionization neon atoms
in the Galaxy
would not likely be responsible for any absorption near 13.8\AA\ in the spectrum of \xc.

\section{Conclusions}

\xmm\ observations of \xc\ with the \rgs\ spectrometers, combined
with measurements of the soft X--ray emission from \rosat, have provided
useful constraints to the properteis of the WHIM near the
\coma\ cluster. First, we have searched for the presence
of absorption lines from several ions
that are expected to be abundant at $\log T(K)=6-7$, following
the earlier study of \cite{takei2007} based on a sub--set
of these observations. We do not
find significant absorption from any of the prominent
X--ray ions.
We do find an absorption signal
that is consistent with 
 \neix\ at the redshift of
\coma, statistically significant between the 90\% and 99\% confidence level. 
It is therefore in principle possible that 
there are significant amounts of warm--hot absorbing plasma along this sightline,
but our analysis cannot confirm their presence with high statistical significance.
%At the same redshift, there is only a marginal deficit of 
%photons at the wavelengths
%of the redshifted \oviii\ line, relative to the source's continuum. In our analysis %we %therefore
%find a similar significance of detection for putative
%WHIM lines towards \coma\ as in \cite{takei2007} 
%At the wavelengths of the redshifted \neix\ line, only one of the two \rgs\ cameras has
%useful data, and no observations with \chandra\ have been performed
%towards \xc. It would therefore be necessary to confirm these measurements
%with higher--resolution \xmm\ and \chandra\ data.

\neix, \oviii\ and \ovii\ are the most prominent ions 
for a WHIM 
 temperature of $kT \simeq \nicefrac{1}{4}$~keV, the temperature  indicated
by the \cite{bonamente2003} observations
of the soft X--ray emission towards \coma.
We have therefore combined 
 the \rgs\  absorption line measurements with \rosat\
measurements of the soft X--ray emission from
\coma\ \citep{bonamente2003} to study the overall characteristics
of the WHIM towards \coma.
We used the measured upper limits 
to place constraints on the density and the length along the sightline of the WHIM filaments, and
find that the WHIM towards \coma\ has properties that are
typical for the WHIM, with a sightline distance of less than a few hundred Mpc
and with a baryon overdensity in excess of approximately 10. 
The soft X--ray emission and absorption line spectroscopy presented
in this paper are generally consistent with the scenario that \coma\ may host a WHIM that
converges to the cluster from nearby filaments. 
Our search for galaxy filaments
with \sdss\ data however proved inconclusive.
We showed that the galaxy confusion does not take place, i.e. that galactic 
halos along the  sightline towards \xc\ would not be able to produce an absorption at the level
of the measured upper limits.

Our analysis of the physical properties of WHIM towards the 
\xc\ sight line is necessarily driven by the available observations. 
Limited by marginal constraints on only a few ionic species, 
our model assumes a single-phase gas, but cosmological simulations suggest 
that multi-temperature structure is likely \citep[e.g.,][]{yoshida2005}. 
Addressing these limitations and  leveraging \xc's  position behind the 
\coma\ cluster will require further observations. 
Independent \chandra\ observations could confirm and further 
constrain the X-ray transitions discussed above. Additionally,
simulations predict substantial gas around \coma-like clusters in 
phases detectable via FUV transitions \citep[e.g.,][]{emerick2015}. 
As \xc\ has only been observed in the FUV with Hubble Space Telescope's 
(\hst) FOS/G130H at very low signal-to-noise insufficient for detailed 
absorption line studies, new  observations with the 
Cosmic Origins Spectrograph on \hst\ could directly confirm multi-temperature 
WHIM along the line of sight. \xc's unique position on the sky offers 
a unique  probe of the WHIM near galaxy clusters, 
which can only be fully exploited if such additional observations 
are undertaken.

Acknowledgements:
%J. Nevalainen  is  funded  by  PUT246  grant  from Estonian  Research Council.
We thank prof. A. Finoguenov for help in the interpretation of the data.
We thank L. Liivam\"{a}gi %, P. Hein\"{a}m\"{a}ki
for help in the analysis of the data. TF was partially supported by the
National Natural Science Foundation of China under grant No.~11273021
E. Tempel acknowledges the support by the ETAg grants IUT26-2, IUT40-2, 
and by the European Regional Development Fund (TK133).
E. Tilton acknowledges support from NASA Earth and Space Science Fellowship grant NNX14-AO18H

\bibliographystyle{mn2e}
\bibliography{/home/max/proposals/max}

\begin{thebibliography}{48}
\expandafter\ifx\csname natexlab\endcsname\relax\def\natexlab#1{#1}\fi

\bibitem[{{Anders} \& {Grevesse}(1989)}]{anders1989}
{Anders} E., {Grevesse} N., 1989, \gca, 53, 197

\bibitem[{{Arnaud} {et~al}\mbox{.}(2001){Arnaud}, {Aghanim}, {Gastaud},
  {Neumann}, {Lumb}, {Briel}, {Altieri}, {Ghizzardi}, {Mittaz}, {Sasseen}, \&
  {Vestrand}}]{arnaud2001}
{Arnaud} M. {et~al.}, 2001, \aap, 365, L67

\bibitem[{{Bonamente} {et~al}\mbox{.}(2003){Bonamente}, {Joy}, \&
  {Lieu}}]{bonamente2003}
{Bonamente} M., {Joy} M.~K., {Lieu} R., 2003, \apj, 585, 722

\bibitem[{{Bonamente} {et~al}\mbox{.}(2009){Bonamente}, {Lieu}, \&
  {Bulbul}}]{bonamente2009}
{Bonamente} M., {Lieu} R., {Bulbul} E., 2009, \apj, 696, 1886

\bibitem[{{Bonamente} {et~al}\mbox{.}(2002){Bonamente}, {Lieu}, {Joy}, \&
  {Nevalainen}}]{bonamente2002}
{Bonamente} M., {Lieu} R., {Joy} M.~K., {Nevalainen} J.~H., 2002, \apj, 576,
  688

\bibitem[{{Bonamente} {et~al}\mbox{.}(2016){Bonamente}, {Nevalainen}, {Tilton},
  {Liivam{\"a}gi}, {Tempel}, {Hein{\"a}m{\"a}ki}, \& {Fang}}]{bonamente2016}
{Bonamente} M., {Nevalainen} J., {Tilton} E., {Liivam{\"a}gi} J., {Tempel} E.,
  {Hein{\"a}m{\"a}ki} P., {Fang} T., 2016, \mnras, 457, 4236

\bibitem[{{Branduardi-Raymont} {et~al}\mbox{.}(1985){Branduardi-Raymont},
  {Mason}, {Murdin}, \& {Martin}}]{Branduardi1985}
{Branduardi-Raymont} G., {Mason} K.~O., {Murdin} P.~G., {Martin} C., 1985,
  \mnras, 216, 1043

\bibitem[{{Cash}(1979)}]{cash1979}
{Cash} W., 1979, \apj, 228, 939

\bibitem[{{Cen} \& {Ostriker}(1999)}]{cen1999}
{Cen} R., {Ostriker} J.~P., 1999, \apj, 514, 1

\bibitem[{{Danforth} {et~al}\mbox{.}(2016){Danforth}, {Keeney}, {Tilton},
  {Shull}, {Stocke}, {Stevans}, {Pieri}, {Savage}, {France}, {Syphers},
  {Smith}, {Green}, {Froning}, {Penton}, \& {Osterman}}]{danforth2016}
{Danforth} C.~W. {et~al.}, 2016, \apj, 817, 111

\bibitem[{{Danforth} \& {Shull}(2008)}]{danforth2008}
{Danforth} C.~W., {Shull} J.~M., 2008, \apj, 679, 194

\bibitem[{{Dav{\'e}} {et~al}\mbox{.}(2001){Dav{\'e}}, {Cen}, {Ostriker},
  {Bryan}, {Hernquist}, {Katz}, {Weinberg}, {Norman}, \& {O'Shea}}]{dave2001}
{Dav{\'e}} R. {et~al.}, 2001, \apj, 552, 473

\bibitem[{{Emerick} {et~al}\mbox{.}(2015){Emerick}, {Bryan}, \&
  {Putman}}]{emerick2015}
{Emerick} A., {Bryan} G., {Putman} M.~E., 2015, \mnras, 453, 4051

\bibitem[{{Finoguenov} {et~al}\mbox{.}(2003){Finoguenov}, {Briel}, \&
  {Henry}}]{finoguenov2003}
{Finoguenov} A., {Briel} U.~G., {Henry} J.~P., 2003, \aap, 410, 777

\bibitem[{{Fukazawa} {et~al}\mbox{.}(2006){Fukazawa}, {Botoya-Nonesa}, {Pu},
  {Ohto}, \& {Kawano}}]{fukazawa2006}
{Fukazawa} Y., {Botoya-Nonesa} J.~G., {Pu} J., {Ohto} A., {Kawano} N., 2006,
  \apj, 636, 698

\bibitem[{{Gatuzz} {et~al}\mbox{.}(2015){Gatuzz}, {Garc{\'{\i}}a}, {Kallman},
  {Mendoza}, \& {Gorczyca}}]{gatuzz2015}
{Gatuzz} E., {Garc{\'{\i}}a} J., {Kallman} T.~R., {Mendoza} C., {Gorczyca}
  T.~W., 2015, \apj, 800, 29

\bibitem[{{Gatuzz} {et~al}\mbox{.}(2013){Gatuzz}, {Garc{\'{\i}}a}, {Mendoza},
  {Kallman}, {Witthoeft}, {Lohfink}, {Bautista}, {Palmeri}, \&
  {Quinet}}]{gatuzz2013}
{Gatuzz} E. {et~al.}, 2013, \apj, 768, 60

\bibitem[{{Geller} {et~al}\mbox{.}(1999){Geller}, {Diaferio}, \&
  {Kurtz}}]{geller1999}
{Geller} M.~J., {Diaferio} A., {Kurtz} M.~J., 1999, \apjl, 517, L23

\bibitem[{{Gnat} \& {Sternberg}(2007)}]{gnat2007}
{Gnat} O., {Sternberg} A., 2007, \apjs, 168, 213

\bibitem[{{Hughes} {et~al}\mbox{.}(1993){Hughes}, {Butcher}, {Stewart}, \&
  {Tanaka}}]{hughes1993}
{Hughes} J.~P., {Butcher} J.~A., {Stewart} G.~C., {Tanaka} Y., 1993, \apj, 404,
  611

\bibitem[{{Jackson}(1972)}]{jackson1972}
{Jackson} J.~C., 1972, \mnras, 156, 1P

\bibitem[{{Kaastra} {et~al}\mbox{.}(1996){Kaastra}, {Mewe}, \&
  {Nieuwenhuijzen}}]{kaastra1996}
{Kaastra} J.~S., {Mewe} R., {Nieuwenhuijzen} H., 1996, in UV and X-ray
  Spectroscopy of Astrophysical and Laboratory Plasmas, {Yamashita} K.,
  {Watanabe} T., eds., pp. 411--414

\bibitem[{{Kim} {et~al}\mbox{.}(2013){Kim}, {Choi}, \& {Kim}}]{kim2013}
{Kim} E., {Choi} Y.-Y., {Kim} S.~S., 2013, Journal of Korean Astronomical
  Society, 46, 33

\bibitem[{{Kubo} {et~al}\mbox{.}(2007){Kubo}, {Stebbins}, {Annis},
  {Dell'Antonio}, {Lin}, {Khiabanian}, \& {Frieman}}]{kubo2007}
{Kubo} J.~M., {Stebbins} A., {Annis} J., {Dell'Antonio} I.~P., {Lin} H.,
  {Khiabanian} H., {Frieman} J.~A., 2007, \apj, 671, 1466

\bibitem[{{Lieu} {et~al}\mbox{.}(1996){Lieu}, {Mittaz}, {Bowyer}, {Breen},
  {Lockman}, {Murphy}, \& {Hwang}}]{lieu1996b}
{Lieu} R., {Mittaz} J.~P.~D., {Bowyer} S., {Breen} J.~O., {Lockman} F.~J.,
  {Murphy} E.~M., {Hwang} C.-Y., 1996, Science, 274, 1335

\bibitem[{{Mazzotta} {et~al}\mbox{.}(1998){Mazzotta}, {Mazzitelli},
  {Colafrancesco}, \& {Vittorio}}]{mazzotta1998}
{Mazzotta} P., {Mazzitelli} G., {Colafrancesco} S., {Vittorio} N., 1998, \aaps,
  133, 403

\bibitem[{{M{\"u}ller} {et~al}\mbox{.}(2017){M{\"u}ller}, {Bernhardt},
  {Borovik}, {Buhr}, {Hellhund}, {Holste}, {Kilcoyne}, {Klumpp}, {Martins},
  {Ricz}, {Seltmann}, {Viefhaus}, \& {Schippers}}]{muller2017}
{M{\"u}ller} A. {et~al.}, 2017, \apj, 836, 166

\bibitem[{{Nevalainen} {et~al}\mbox{.}(2015){Nevalainen}, {Tempel},
  {Liivam{\"a}gi}, {Branchini}, {Roncarelli}, {Giocoli}, {Hein{\"a}m{\"a}ki},
  {Saar}, {Tamm}, {Finoguenov}, {Nurmi}, \& {Bonamente}}]{nevalainen2015}
{Nevalainen} J. {et~al.}, 2015, \aap, 583, A142

\bibitem[{{Nicastro} {et~al}\mbox{.}(2016{\natexlab{a}}){Nicastro}, {Senatore},
  {Gupta}, {Guainazzi}, {Mathur}, {Krongold}, {Elvis}, \&
  {Piro}}]{nicastro2016}
{Nicastro} F., {Senatore} F., {Gupta} A., {Guainazzi} M., {Mathur} S.,
  {Krongold} Y., {Elvis} M., {Piro} L., 2016{\natexlab{a}}, \mnras, 457, 676

\bibitem[{{Nicastro} {et~al}\mbox{.}(2016{\natexlab{b}}){Nicastro}, {Senatore},
  {Gupta}, {Mathur}, {Krongold}, {Elvis}, \& {Piro}}]{nicastro2016b}
{Nicastro} F., {Senatore} F., {Gupta} A., {Mathur} S., {Krongold} Y., {Elvis}
  M., {Piro} L., 2016{\natexlab{b}}, \mnras, 458, L123

\bibitem[{{Penton} {et~al}\mbox{.}(2000){Penton}, {Shull}, \&
  {Stocke}}]{penton2000}
{Penton} S.~V., {Shull} J.~M., {Stocke} J.~T., 2000, \apj, 544, 150

\bibitem[{{Planck Collaboration} {et~al}\mbox{.}(2013){Planck Collaboration},
  {Ade}, {Aghanim}, {Arnaud}, {Ashdown}, {Atrio-Barandela}, {Aumont},
  {Baccigalupi}, {Balbi}, {Banday}, \& et~al.}]{planck2013-coma}
{Planck Collaboration} {et~al.}, 2013, \aap, 554, A140

\bibitem[{{Schmidt} {et~al}\mbox{.}(2016){Schmidt}, {Engels}, {Niemeyer}, \&
  {Almgren}}]{schmidt2016}
{Schmidt} W., {Engels} J.~F., {Niemeyer} J.~C., {Almgren} A.~S., 2016, \mnras,
  459, 701

\bibitem[{{Shull} {et~al}\mbox{.}(2012){Shull}, {Smith}, \&
  {Danforth}}]{shull2012}
{Shull} J.~M., {Smith} B.~D., {Danforth} C.~W., 2012, \apj, 759, 23

\bibitem[{{Simionescu} {et~al}\mbox{.}(2013){Simionescu}, {Werner}, {Urban},
  {Allen}, {Fabian}, {Mantz}, {Matsushita}, {Nulsen}, {Sanders}, {Sasaki},
  {Sato}, {Takei}, \& {Walker}}]{simionescu2013}
{Simionescu} A. {et~al.}, 2013, \apj, 775, 4

\bibitem[{{Snowden} {et~al}\mbox{.}(1994){Snowden}, {McCammon}, {Burrows}, \&
  {Mendenhall}}]{snowden1994}
{Snowden} S.~L., {McCammon} D., {Burrows} D.~N., {Mendenhall} J.~A., 1994,
  \apj, 424, 714

\bibitem[{{Struble} \& {Rood}(1999)}]{struble1999}
{Struble} M.~F., {Rood} H.~J., 1999, \apjs, 125, 35

\bibitem[{{Takei} {et~al}\mbox{.}(2007){Takei}, {Henry}, {Finoguenov},
  {Mitsuda}, {Tamura}, {Fujimoto}, \& {Briel}}]{takei2007}
{Takei} Y., {Henry} J.~P., {Finoguenov} A., {Mitsuda} K., {Tamura} T.,
  {Fujimoto} R., {Briel} U.~G., 2007, \apj, 655, 831

\bibitem[{{Takei} {et~al}\mbox{.}(2008){Takei}, {Miller}, {Bregman}, {Kimura},
  {Ohashi}, {Mitsuda}, {Tamura}, {Yamasaki}, \& {Fujimoto}}]{takei2008}
{Takei} Y. {et~al.}, 2008, \apj, 680, 1049

\bibitem[{{Tempel} {et~al}\mbox{.}(2016){Tempel}, {Kipper}, {Tamm}, {Gramann},
  {Einasto}, {Sepp}, \& {Tuvikene}}]{tempel2016}
{Tempel} E., {Kipper} R., {Tamm} A., {Gramann} M., {Einasto} M., {Sepp} T.,
  {Tuvikene} T., 2016, \aap, 588, A14

\bibitem[{{Tempel} {et~al}\mbox{.}(2014{\natexlab{a}}){Tempel}, {Stoica},
  {Mart{\'{\i}}nez}, {Liivam{\"a}gi}, {Castellan}, \& {Saar}}]{tempel2014a}
{Tempel} E., {Stoica} R.~S., {Mart{\'{\i}}nez} V.~J., {Liivam{\"a}gi} L.~J.,
  {Castellan} G., {Saar} E., 2014{\natexlab{a}}, \mnras, 438, 3465

\bibitem[{{Tempel} {et~al}\mbox{.}(2014{\natexlab{b}}){Tempel}, {Tamm},
  {Gramann}, {Tuvikene}, {Liivam{\"a}gi}, {Suhhonenko}, {Kipper}, {Einasto}, \&
  {Saar}}]{tempel2014b}
{Tempel} E. {et~al.}, 2014{\natexlab{b}}, \aap, 566, A1

\bibitem[{{Tilton} {et~al}\mbox{.}(2012){Tilton}, {Danforth}, {Shull}, \&
  {Ross}}]{tilton2012}
{Tilton} E.~M., {Danforth} C.~W., {Shull} J.~M., {Ross} T.~L., 2012, \apj, 759,
  112

\bibitem[{{Tripp} {et~al}\mbox{.}(2008){Tripp}, {Sembach}, {Bowen}, {Savage},
  {Jenkins}, {Lehner}, \& {Richter}}]{tripp2008}
{Tripp} T.~M., {Sembach} K.~R., {Bowen} D.~V., {Savage} B.~D., {Jenkins} E.~B.,
  {Lehner} N., {Richter} P., 2008, \apjs, 177, 39

\bibitem[{{Tully} \& {Fisher}(1978)}]{tully1978}
{Tully} R.~B., {Fisher} J.~R., 1978, in IAU Symposium, Vol.~79, Large Scale
  Structures in the Universe, {Longair} M.~S., {Einasto} J., eds., pp. 31--45

\bibitem[{{Verner} {et~al}\mbox{.}(1996){Verner}, {Verner}, \&
  {Ferland}}]{verner1996}
{Verner} D.~A., {Verner} E.~M., {Ferland} G.~J., 1996, Atomic Data and Nuclear
  Data Tables, 64, 1

\bibitem[{{Werner} {et~al}\mbox{.}(2008){Werner}, {Finoguenov}, {Kaastra},
  {Simionescu}, {Dietrich}, {Vink}, \& {B{\"o}hringer}}]{werner2008}
{Werner} N., {Finoguenov} A., {Kaastra} J.~S., {Simionescu} A., {Dietrich}
  J.~P., {Vink} J., {B{\"o}hringer} H., 2008, \aap, 482, L29

\bibitem[{{Yoshida} {et~al}\mbox{.}(2005){Yoshida}, {Furlanetto}, \&
  {Hernquist}}]{yoshida2005}
{Yoshida} N., {Furlanetto} S.~R., {Hernquist} L., 2005, \apjl, 618, L91

\end{thebibliography}

\end{document}